%% file: main_ndss18.tex
\newcommand{\code}[1]{\texttt{\small #1}}
\documentclass[conference]{IEEEtran}

\usepackage{pgfplots}
\usepackage{bchart}
\usepackage{caption}
\usepackage{array}
\newcolumntype{P}[1]{>{\centering\arraybackslash}p{#1}}
\usepackage{xcolor}
\usepackage{fancyhdr}

\fancypagestyle{firstpage}{%
  \chead{{\color{blue} {\large To appear at the 2018 Network and Distributed System Security Symposium (NDSS).}}}
  \rhead{}
}
\ifCLASSINFOpdf
\else
\fi

\begin{document}

\title{{\color{blue} {\large To appear at the 2018 Network and Distributed System Security Symposium (NDSS).}}\\When Coding Style Survives Compilation:\\De-anonymizing Programmers from Executable Binaries}

%\author{ 
%\IEEEauthorblockN{ 
%Aylin Caliskan}
%       \IEEEauthorblockA{Princeton University\\
%       aylinc@princeton.edu}
%\and
%\IEEEauthorblockN{  
%Fabian Yamaguchi}
%       \IEEEauthorblockA{ShiftLeft Inc\\
%       fabs@shiftleft.io}
%      \and       
% \IEEEauthorblockN{ 
% Edwin Dauber}
%       \IEEEauthorblockA{Drexel University\\
%       egd34@drexel.edu}
%       \and
%\IEEEauthorblockN{ 
%Konrad Rieck}
%       \IEEEauthorblockA{TU Braunschweig\\
%       k.rieck@tu-bs.de}
%       \and       
%\IEEEauthorblockN{ 
%Richard Harang}
%       \IEEEauthorblockA{Sophos, Data Science Team\\
%      rich.harang@sophos.com}
%\and
%\IEEEauthorblockN{ 
%Rachel Greenstadt} 
%       \IEEEauthorblockA{Drexel University\\
%        rachel.a.greenstadt@drexel.edu}
%       \and
%\IEEEauthorblockN{Arvind Narayanan}
%       \IEEEauthorblockA{Princeton University\\
%       arvindn@cs.princeton.edu}
%}

\author{
  \IEEEauthorblockN{
    Aylin Caliskan\IEEEauthorrefmark{1}, 
    Fabian Yamaguchi\IEEEauthorrefmark{2}, 
    Edwin Dauber\IEEEauthorrefmark{3}, \\
   Richard Harang\IEEEauthorrefmark{4},
   Konrad Rieck\IEEEauthorrefmark{5}, 
      Rachel Greenstadt\IEEEauthorrefmark{3} and
          Arvind Narayanan\IEEEauthorrefmark{1}}
          \IEEEauthorblockA{\IEEEauthorrefmark{1}Princeton University, \{aylinc@, arvindn@cs\}.princeton.edu}
    \IEEEauthorblockA{\IEEEauthorrefmark{2}Shiftleft Inc, fabs@shiftleft.io}
    \IEEEauthorblockA{\IEEEauthorrefmark{3}Drexel University, \{egd34, rachel.a.greenstadt\}@drexel.edu}
     \IEEEauthorblockA{\IEEEauthorrefmark{4}Sophos, Data Science Team, rich.harang@sophos.com}
    \IEEEauthorblockA{\IEEEauthorrefmark{5}TU Braunschweig, k.rieck@tu-bs.de}}

\IEEEoverridecommandlockouts
\makeatletter\def\@IEEEpubidpullup{9\baselineskip}\makeatother
\IEEEpubid{\parbox{\columnwidth}{
    Network and Distributed Systems Security (NDSS) Symposium 2018\\
    18-21 February 2018, San Diego, CA, USA\\
    ISBN 1-1891562-49-5\\
    http://dx.doi.org/10.14722/ndss.2018.23304\\
    www.ndss-symposium.org
}
\hspace{\columnsep}\makebox[\columnwidth]{}}

\maketitle

\input{0_abstract}

\input{1_intro}

\input{2_problem}
\input{3_related}

\input{5_approach}
\input{6_experiments}

%\input{7_results}
\input{8_discussion}
\input{9_conclusion}

% Explain why obfuscations do not work and refer to semantics aware malware detection
% Explain the normalizations used in features
% explain that we actually have less features if the regex was more specific
%now we have equivalent features because of the regex
% add new features, skip grams and longer grams to handle obfuscations, and weigh the features in case too much noise is inserted.

%\section*{Acknowledgment}
%This material is based on work supported by the ARO (U.S. Army Research Office) Grant W911NF-15-2-0055 and AWS in Education Research Grant award. The views and conclusions contained herein are those of the authors and should not be interpreted as representing the official policies, either expressed or implied, of the Army Research Laboratory or the U.S. Government. The U.S. Government is authorized to reproduce and distribute reprints for Government purposes notwithstanding any copyright notice herein. This material is based on work supported by the ARO (U.S. Army Research Office) Grant W911NF-14-1-0444, the DFG (German Research Foundation) under the project DEVIL (RI 2469/1- 1), and AWS in Education Research Grant award. This research was supported in part by the Center for Information Technology Policy at Princeton University.

%{\footnotesize \bibliographystyle{acm}
%FIX THIS!

\bibliographystyle{IEEEtranS}
\bibliography{ndss18}
%\include{appendix}
%\theendnotes

\end{document}

%% file: 0_abstract.tex
\begin{abstract}
%The ability to identify authors of programs based on their style is a direct threat for the privacy of anonymous code contributors. In addition, it aids in forensic analysis by hinting towards possible authors of malicious code left on compromised systems. Previous work in the area shows that de-anonymizing programmers from source code is often possible with high accuracy, however, whether effective methods for authorship attribution exist for executable binary is currently unknown.
%%
%In this work, we investigate methods to de-anonymize authors of executable binary using stylistic fingerprints. We cast binary authorship attribution as a machine learning problem using features obtained from disassemblers and decompilers. We thus obtain a unique feature set, that serves as a novel representation of stylistic properties of code that remain recoverable from executable binary. Our result binary de-anonymization method based on random forest classifiers attributes more authors (100) with significantly higher accuracy (78\%) on a larger data set (Google Code Jam) than has been previously attempted. Furthermore our method is more robust than previous methods, allowing authors to be attributed even for code compiled with different levels of optimization.
%

The ability to identify authors of computer programs based on their coding style is a direct threat to the privacy and anonymity of programmers. While recent work found that source code can be attributed to authors with high accuracy, attribution of executable binaries appears to be much more difficult. Many distinguishing features present in source code, e.g. variable names, are removed in the compilation process, and compiler optimization may alter the structure of a program, further obscuring features that are known to be useful in determining authorship. We examine programmer de-anonymization from the standpoint of machine learning, using a novel set of features that include ones obtained by decompiling the executable binary to source code. We adapt a powerful set of techniques from the domain of source code authorship attribution along with stylistic representations embedded in assembly, resulting in successful de-anonymization of a large set of programmers. 

We evaluate our approach on data from the Google Code Jam, obtaining attribution accuracy of up to 96\% with 100 and 83\% with 600 candidate programmers. We present an executable binary authorship attribution approach, for the first time, that is robust to basic obfuscations, a range of compiler optimization settings, and binaries that have been stripped of their symbol tables. We perform programmer de-anonymization using both obfuscated binaries, and real-world code found ``in the wild'' in single-author GitHub repositories and the recently leaked Nulled.IO hacker forum. We show that programmers who would like to remain anonymous need to take extreme countermeasures to protect their privacy.

%\begin{IEEEkeywords}
% Authorship attribution; Decompilation; Machine Learning;
%\end{IEEEkeywords}

%This analysis also produces interesting insights relevant to software engineering. We find that (i) the code resulting from difficult programming tasks is easier to attribute than easier tasks and (ii) skilled programmers (who can complete the more difficult tasks) are easier to attribute than less skilled programmers.

\end{abstract}

%% file: 1_intro.tex
\section{Introduction}
\label{sec:introduction}

% Introduction to the problem

If we encounter an executable binary sample in the wild, what can we learn from it? In this work, we show that the programmer's stylistic fingerprint, or coding style, is preserved in the compilation process and can be extracted from the executable binary. This means that it may be possible to infer the programmer's identity if we have a set of known potential candidate programmers, along with executable binary samples (or source code) known to be authored by these candidates.

Programmer de-anonymization from executable binaries has implications for privacy and anonymity. Perhaps the creator of a censorship circumvention tool distributes it anonymously, fearing repression. Our work shows that such a programmer might be de-anonymized. Further, there are applications for software forensics, for example to help adjudicate cases of disputed authorship or copyright.

The White House Cyber R\&D Plan states that ``effective deterrence must raise the cost of malicious cyber activities, lower their 
gains, and convince adversaries that such activities can be attributed~\cite{cyberrd}.'' The DARPA Enhanced Attribution calls for methods that can ``consistently identify virtual personas and individual malicious cyber operators over time and across different endpoint devices and C2 infrastructures~\cite{ea-baa}.'' While the forensic applications are important, as attribution methods develop, they will threaten the anonymity of privacy-minded individuals at least as much as malicious actors.

We introduce the first part of our approach by significantly overperforming the previous attempt at de-anonymizing programmers by Rosenblum et al.~\cite{rosenblum2011wrote}. We improve their accuracy of 51\% in de-anonymizing 191 programmers to 92\% and then we scale the results to 83\% accuracy on 600 programmers. First, whereas Rosenblum et al. extract structures such as control-flow graphs directly from the executable binaries, our work is the first to show that {\em automated decompilation} of executable binaries gives additional categories of useful features. Specifically, we generate {\em abstract syntax trees} of decompiled source code. Abstract syntax trees have been shown to greatly improve author attribution of source code~\cite{caliskan2015anonymizing}. We find that syntactical properties derived from these trees also improve the accuracy of executable binary attribution techniques.

Second, we demonstrate that using multiple tools for disassembly and decompilation in parallel increases the accuracy of de-anonymization by generating different representations of code that capture various aspects of the programmer's style. We present a robust machine learning framework based on entropy and correlation for dimensionality reduction, followed by random-forest classification, that allows us to effectively use disparate types of features in conjunction without overfitting.

These innovations allow us to de-anonymize a large set of real-world programmers with high accuracy. We perform experiments with a controlled dataset collected from Google Code Jam (GCJ), allowing a direct comparison to previous work that used samples from GCJ. The results of these experiments are discussed in detail in Section~\ref{sec:experiments}. Specifically; we can distinguish between {\em thirty times} as many candidate programmers (600 vs. 20) with higher accuracy, while utilizing less training data and  much fewer stylistic features (53) per programmer. The accuracy of our method degrades gracefully as the number of programmers increases, and we present experiments with as many as 600 programmers. Similarly, we are able to tolerate scarcity of training data: our accuracy for de-anonymizing sets of 20 candidate programmers with just a single training sample per programmer is over 75\%.

Third, we find that traditional binary obfuscation, enabling compiler optimizations, or stripping debugging symbols in executable binaries results in only a modest decrease in de-anonymization accuracy. These results, described in Section~\ref{sec:compExp}, are an important step toward establishing the practical significance of the method.

The fact that coding style survives compilation is unintuitive, and may leave the reader wanting a ``sanity check" or an explanation for why this is possible. In Section~\ref{subsec:reconstruction}, we present several experiments that help illuminate this mystery. First, we show that decompiled source code {\em is not} necessarily similar to the original source code in terms of the features that we use; rather, the feature vector obtained from disassembly and decompilation can be used to {\em predict}, using machine learning, the features in the original source code. Even if no individual feature is well preserved, there is enough information in the vector as a whole to enable this prediction. On average, the cosine similarity between the original feature vector and the reconstructed vector is over 80\%. Further, we investigate factors that are correlated with coding style being well-preserved, and find that more skilled programmers are more fingerprintable. This suggests that programmers gradually acquire their own unique style as they gain experience.

All these experiments were carried out using the GCJ dataset; the availability of this dataset is a boon for research in this area since it allows us to develop and benchmark our results under controlled settings \cite{rosenblum2011wrote, alrabaee2014oba2}. Having done that, we present the first ever de-anonymization study on an uncontrolled real-world dataset collected from GitHub in Section~\ref{sec:gitExp}. This data presents difficulties, particularly noise in ground truth because of library and code reuse. However, we show that we can handle a noisy dataset of 50 programmers found in the wild with 65\% accuracy and further extend our method to tackle open world scenarios. We also present a case study using code found via the recently leaked Nulled.IO hacker forum. We were able to find four forum members who, in private messages, linked to executables they had authored (one of which had only one sample). Our approach correctly attributed the three individuals who had enough data to build a model and correctly rejected the fourth sample as none of the previous three. 

We emphasize that research challenges remain before programmer de-anonymization from executable binaries is fully ready for practical use. For example, programs may be authored by multiple programmers and may have gone through encryption. We have not performed experiments that model these scenarios which require different machine learning and segmentation techniques and we mainly focus on the privacy implications. Nonetheless, we present a robust and principled programmer de-anonymization method with a new approach and for the first time explore various realistic scenarios. Accordingly, our effective framework raise immediate concerns for privacy and anonymity.

The remainder of this paper is structured as follows. We begin by formulating the research question investigated throughout this paper in Section~\ref{sec:problem}, and discuss closely related work on de-anonymization in Section~\ref{sec:related}.  We proceed to describe our novel approach for binary authorship attribution based on instruction information, control flow graphs, and decompiled code in Section~\ref{sec:approach}. Our experimental results are described in Section~\ref{sec:experiments}, followed by a discussion of results in Section~\ref{sec:discussion}. Finally, we shed light on the limitations of our method in Section~\ref{sec:limitations} and conclude in Section~\ref{sec:conclusion}.
%
%
%
%4. in the introduction, maybe move the accuracy to the end. that's not the most important contribution
%5. in general, you should express our surprise that source code features are preserved in the binary at all
%b)	Contributions
%	i)	improvement in accuracy: rephrase this, instead of focusing on improvement in accuracy, explain how this introduces a robust method (in this domain, this is a breakthrough in accuracy-not an incremental improvement)
%	ii)	new features
%(1)	from our own disassembler-bjoern

%% file: 2_problem.tex
\section{Problem Statement}
\label{sec:problem}

In this work, we consider an analyst interested in determining the author of an executable binary purely based on its style. Moreover, we assume that the analyst only has access to executable binary samples each assigned to one of a set of candidate programmers. 

Depending on the context, the analyst's goal might be defensive or offensive in nature. For example, the analyst might be trying to identify a misbehaving employee that violates the non-compete clause in his company by launching an application related to what he does at work. 
%Similarly, a malware analyst might be interested in finding the author or authors of a malicious executable binary.
By contrast, the analyst might belong to a surveillance agency in an oppressive regime who tries to unmask anonymous programmers.
The regime might have made it unlawful for its citizens to use certain types of programs, such as censorship-circumvention tools, and might want to punish the programmers of any such tools. 
If executable binary stylometry is possible, it means that compiled and cryptic code does not guarantee anonymity. Because of its potential dual use, executable binary stylometry is of interest to both security and privacy researchers. 

In either (defensive or offensive) case, the analyst (or adversary) will seek to obtain labeled executable binary samples from each of these programmers who may have potentially authored the anonymous executable binary. 
The analyst proceeds by converting each labeled sample into a numerical feature vector, and subsequently deriving a classifier from these vectors using machine learning techniques. This classifier can then be used to attribute the anonymous executable binary to the most likely programmer.

Since we assume that a set of candidate programmers is known, we treat our main problem as a closed world, supervised machine learning task. It is a multi-class machine learning problem where the classifier calculates the most likely author for the anonymous executable binary sample among multiple authors. We also present experiments on an open-world scenario in Section~\ref{sec:openworld}.

\textbf{Stylistic Fingerprints.}
An analyst is interested in identifying stylistic fingerprints in binary code to show that compiling source code does not anonymize it. The analyst engineers the numeric representations of stylistic properties that can be derived from binary code. To do so, the analyst generates representations of the program from the binary code. First, she uses a disassembler to obtain the low level features in assembly code. Second, she uses a decompiler to generate the control flow graph to capture the flow of the program. Lastly, she utilizes a decompiler to convert the low level instructions to high level decompiled source code in order to obtain abstract syntax trees. The analyst uses these three data formats to numerically represent the stylistic properties embedded in binary code. Given a set of labeled binary code samples with known authors, the analyst constructs the numeric representation of each sample. The analyst determines the set of stylistic features by calculating how much entropy each numeric value has in correctly differentiating between authors. She can further analyze how programmers' stylistic properties are preserved in a transformed format after compilation. Consequently, the analyst is able to quantify the level of anonymization and the amount of preserved stylistic fingerprints in binary code that has gone through compilation.

%i)	quantifying coding style that survives the compilation process
%ii)	malware authorship attribution
%b)	Verification - commercial software theft
%c)	Compilation does not obfuscate

{\bf Additional Assumptions.} For our experiments, we assume that we know the compiler used for a given program binary. Previous work has shown that with only 20 executable binary samples per compiler as training data, it is possible to use a
linear Conditional Random Field (CRF) to determine the compiler used with accuracy of 93\% on average \cite{rosenblum2010extracting, lafferty2001conditional}.
Other work has shown that by using pattern matching, library functions can be identified with precision and recall between $0.98$ and $1.00$ based
on each one of three criteria; compiler version, library version, and linux distribution \cite{jacobson2011labeling}.

In addition to knowing the compiler, we assume to know the optimization level used for compilation of the binary.
Past work has shown that toolchain provenance, including compiler family, version, optimization, and source language,
can be identified with a linear CRF with accuracy of 99\% for language, compiler family, and optimization
and with 92\% for compiler version \cite{rosenblum2011recovering}. Based on this success, we make the assumption that these
techniques will be used to identify the toolchain provenance of the executable binaries of interest and that our method will be trained using
the same toolchain.

%More recent work has looked at a stratified approach which, although having lower accuracy, is designed to be used at the function level to enable preprocessing for further tasks including authorship attribution \cite{rahimian2015bincomp}. 

%% file: 3_related.tex
\section{Related Work}
\label{sec:related}

Any domain of creative expression allows authors or creators to develop a unique style, and we might expect that there are algorithmic techniques to identify authors based on their style. This class of techniques is called stylometry. Natural-language stylometry, in particular, is well over a century old~\cite{mendenhall1887characteristic}. Other domains such as source code and music also have stylistic features, especially grammar. Therefore stylometry is applicable to these domains as well, often using strikingly similar techniques~\cite{van2006composer, backer2005musical}.

{\bf Linguistic stylometry.}
The state of the art in linguistic stylometry is dominated by machine-learning techniques~\cite{afroz2012detecting, narayanan2012feasibility,dopp}.
Linguistic stylometry has been applied successfully to security and privacy problems, for example Narayanan et al. used stylometry to identify anonymous bloggers in large datasets, exposing privacy issues~\cite{narayanan2012feasibility}. On the other hand, stylometry has also been used for forensics in underground cyber forums. In these forums, the text consists of a mixture of languages and information about forum products, which makes it more challenging to identify personal writing style. Not only have the forum users been de-anonymized but also their multiple identities across and within forums have been linked through stylometric analysis~\cite{dopp}.

Authors may deliberately try to obfuscate or anonymize their writing style~\cite{brennan2012adversarial, afroz2012detecting, mcdonald2012use}. Brennan et al. show how stylometric authorship attribution can be evaded with adversarial stylometry~\cite{brennan2012adversarial}. They present two ways for adversarial stylometry, namely obfuscating writing style and imitating someone else's writing style. Afroz et al. identify the stylistic changes in a piece of writing that has been obfuscated while McDonald et al. present a method to make writing style modification recommendations to anonymize an undisputed document~\cite{afroz2012detecting, mcdonald2012use}.

{\bf Source code stylometry.} Several authors have applied similar techniques to identify programmers based on source code~\cite{caliskan2015anonymizing, pellin2000using, burrows2014comparing}. Source code authorship attribution has applications in software forensics and plagiarism detection\footnote{Note that popular plagiarism-detection tools such as Moss are not based on stylometry; rather they detect code that may have been copied, possibly with modifications. This is an orthogonal problem~\cite{aiken2005moss}.}.

The features used for machine learning in source code authorship attribution range from simple byte-level~\cite{frantzeskou2006effective} and word-level n-grams~\cite{burrows2007source,burrows_info} to more evolved structural features obtained from abstract syntax trees~\cite{caliskan2015anonymizing, pellin2000using}. In particular, Burrows et al. present an approach based on n-grams that reaches an accuracy of 77\% in differentiating 10 different programmers~\cite{burrows_info}.

Similarly, Kothari et al. combine n-grams with lexical markers such as the line length, to build programmer profiles that allow them to identify 12 authors with an accuracy of 76\%~\cite{kothari2007probabilistic}. Lange et al. further show that metrics based on layout and lexical features along with a genetic algorithm reach an accuracy of 75\% in de-anonymizing 20 authors~\cite{lange2007using}. Finally, Caliskan-Islam et al. incorporate abstract syntax tree based structural features to represent programmers' coding style~\cite{caliskan2015anonymizing}. They reach 94\% accuracy in identifying 1,600 programmers of the GCJ data set.

{\bf Executable binary stylometry.} In contrast, identifying programmers from compiled code is considerably more difficult and has received little attention to date.
%In this section, we give an overview of the work on de-anonymization that serves as an inspiration to this work, along with methods for programmer identification that precede our work.
Code compilation results in a loss of information and obstructs stylistic features. We are aware of only two prior works, both of which perform their evaluation and experiments on controlled corpora that are not noisy, such as the GCJ dataset and student homework assignments~\cite{rosenblum2011wrote, alrabaee2014oba2}. Our work significantly overperforms previous work by using different methods and in addition we investigate noisy real-world datasets, an open-world setting, effects of optimizations, and obfuscations.

 \cite{alrabaee2014oba2} present an onion approach for binary code authorship attribution. \cite{rosenblum2011wrote} identify authors of program binaries. Both Alrabaee et al. and Rosenblum et al. utilize the GCJ corpus.

Rosenblum et al. present two main machine learning tasks based on programmer de-anonymization. One is based on supervised classification with a support vector machine to identify the authors of compiled code~\cite{fan2008liblinear}.
%Their feature set that incorporates assembly instructions and control flow graph nodes is able to correctly differentiate between 10 programmers 81\% of the time and 200 programmers 51\% of the time.
The second machine learning approach they use is based on clustering to group together programs written by the same programmers. They incorporate a distance based similarity metric to differentiate between features related to programmer style to increase clustering accuracy. They use the Paradyn project's Parse API for parsing executable binaries to get the instruction sequences and control flow graphs whereas we use four different resources to parse executable binaries to generate a richer representation. Their dataset consists of submissions from GCJ and homework assignments with skeleton code. 

%\hi{How does this API compare to our method?}.
%The first set of generated features capture all possible properties of code without trying to capture stylistic characteristics. Style related features are later calculated using information gain criteria with ten fold cross validation. The final feature set consists of 1,900 features that represent programming style.
%GCJ programmers have eight to sixteen files and students have four to 7 files. 
%Students collaborated on the homework assignments and the skeleton code was available.

{\bf Malware attribution.} While the analysis of malware is a well developed field, authorship attribution of malware has received much less attention. Stylometry may have a role in this application, and this is a ripe area for future work that requires automated packer and encryption detection along with binary segment and metadata analysis. The difficulty in obtaining ground truth labels for malware samples has led much work in this area to focus on clustering malware in some fashion, and the wide range of obfuscation techniques in common use have led many researchers to focus on dynamic analysis rather than the static features we consider. The work of \cite{MarMarGua15} examines several static features intended to provide credible links between executable malware binary produced by the same authors, however many of these features are specific to malware, such as command and control infrastructure and data exfiltration methods, and the authors note that many must be extracted by hand. In dynamic analysis, the work of \cite{pfeffer2012malware} examines information obtained via both static and dynamic analysis of malware samples to organize code samples into lineages that indicate the order in which samples are derived from each other. \cite{bayer2009scalable} convert detailed execution traces from dynamic analysis into more general behavioral profiles, which are then used to cluster malware into groups with related functionality and activity. Supervised methods are used by \cite{rieck2008learning} to match new instances of malware with previously observed families, again on the basis of dynamic analysis.

%% file: 5_approach.tex
\section{Approach}
\label{sec:approach}

\begin{figure*}[t]
 \centering
 \begin{minipage}[b]{1\linewidth}
 \includegraphics[trim= 0 70 0 10, width=\textwidth]{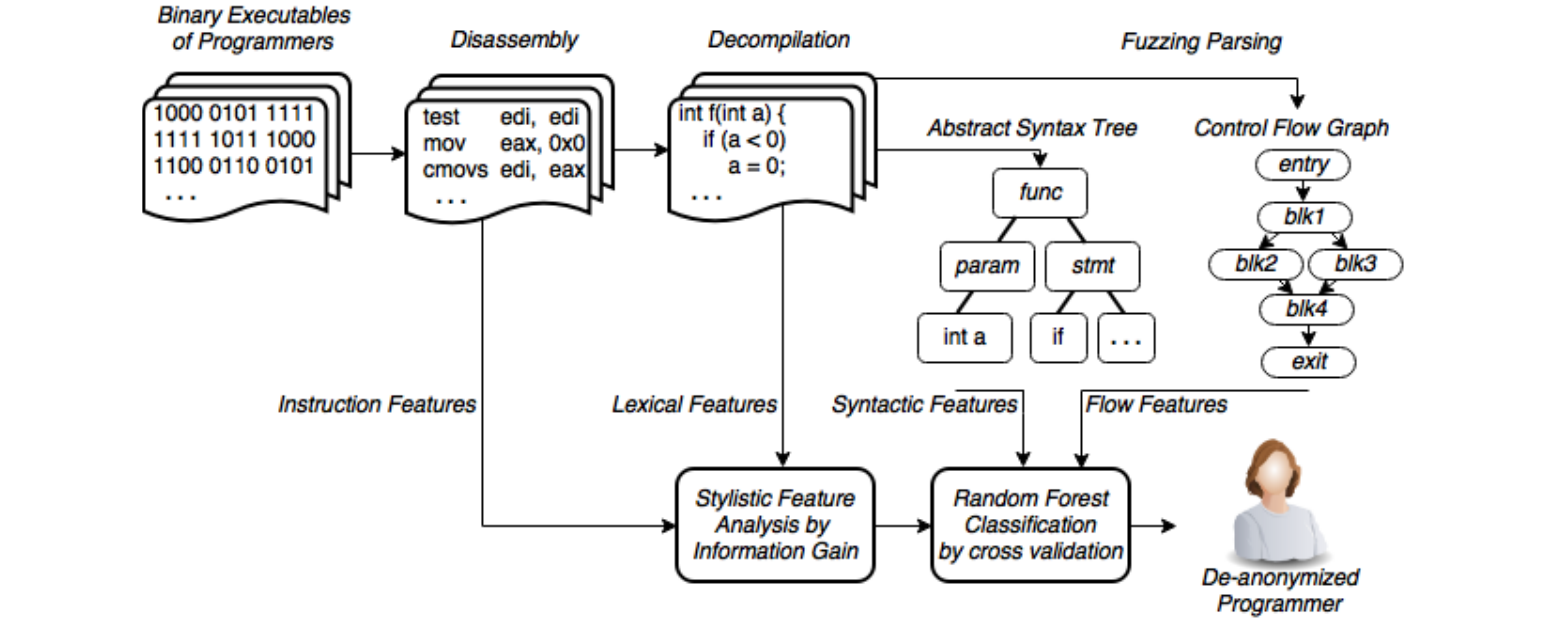}

 \vspace{14mm}

\caption{Overview of our method. Instructions, symbols, and strings are extracted using disassemblers (1), abstract syntax tree and control-flow features are obtained from decompilers (2). Dimensionality reduction first by information gain criteria and then by correlation analysis is performed to obtain features that represent programmer style (3). Finally, a random forest classifier is trained to de-anonymize programmers (4).}
 \label{fig:binflow}
 \end{minipage}
\end{figure*}

Our ultimate goal is to automatically recognize programmers of compiled code. We approach this problem using supervised machine learning, that is, we generate a classifier from training data of sample executable binaries with known authors. The advantage of such learning-based methods over techniques based on manually specified rules is that the approach is easily retargetable to any set of programmers for which sample executable binaries exist. A drawback is that the method is inoperable if samples are not available or too short to represent authorial style. We study the amount of sample data necessary for successful classification in Section~\ref{sec:experiments}.

Data representation is critical to the success of machine learning. Accordingly, we design a feature set for executable binary authorship attribution with the goal of faithfully representing properties of executable binaries relevant for programmer style. We obtain this feature set by augmenting lower-level features extractable from disassemblers with additional string and symbol information, and, most importantly, incorporating higher-level syntactical features obtained from decompilers.

In summary, such an approach results in a method consisting of the following four steps (see Figure~\ref{fig:binflow}) and the code is available at https://github.com/calaylin/bda.

\begin{itemize}
\item \textbf{Disassembly.} We begin by disassembling the program to obtain features based on machine code instructions, referenced strings, symbol information, and control flow graphs (Section~\ref{sec:disassembly}).

\item \textbf{Decompilation.} We proceed to translate the program into C-like pseudo code via decompilation. By subsequently passing the code to a fuzzy parser for C, we thus obtain abstract syntax trees from which syntactical features and n-grams can be extracted (Section~\ref{sec:decompilation}).

\item \textbf{Dimensionality reduction.} With features from disassemblers and decompilers at hand, we select those among them that are particularly useful for classification by employing a standard feature selection technique based on information gain and correlation based feature selection (Section~\ref{sec:dimreduc}).

\item \textbf{Classification.} Finally, a random-forest classifier is trained on the corresponding feature vectors to yield a program that can be used for automatic executable binary authorship attribution (Section~\ref{sec:classification}).
\end{itemize}

In the following sections, we describe these steps in greater detail and provide background information on static code analysis and machine learning where necessary.

\subsection{Feature extraction via disassembly}
\label{sec:disassembly}

As a first step, we disassemble the executable binary to extract low-level features that have been shown to be suitable for authorship attribution in previous work. In particular, we follow the basic example set by Rosenblum et al. and extract raw instruction traces from the executable binary \cite{rosenblum2011wrote}. In addition to this, disassemblers commonly make symbol information available, as well as strings referenced in the code, both of which greatly simplify manual reverse engineering. We augment the feature set accordingly. Finally, we can obtain control flow graphs of functions from disassemblers, providing features based on program basic blocks. The required information necessary to construct our feature set is obtained from the following two disassemblers.

We use two disassemblers to generate two sets of instructions for each binary. We disassemble the binary with the Netwide Disassembler (ndisasm) which is a widely available x86 disassembler. We then use the open source \emph{radare2 disassembler} to get more detailed and higher level instructions than ndisasm's disassembly.

\begin{itemize}

\item \emph{\textbf{The netwide disassembler.}} We begin by exploring whether simple instruction decoding alone can already provide useful features for de-anonymization. To this end, we process each executable binary using the netwide disassembler (\emph{ndisasm}), a rudimentary disassembler that is capable of decoding instructions but is unaware of the executable's file format~\cite{ndisasm15}. Due to this limitation, it resorts to simply decoding the executable binary from start to end, skipping bytes when invalid instructions are encountered. A problem with this approach is that no distinction is made between bytes that represent data versus bytes that represent code. Nonetheless, we explore this simplistic approach as these inaccuracies may not degrade a classifier, given the statistical nature of machine learning.

\item \emph{\textbf{The radare2 disassembler.}} We proceed to apply \emph{radare2}~\cite{radare}, a state-of-the-art open-source disassembler based on the capstone disassembly framework~\cite{capstone15}. In contrast to \emph{ndisasm}, \emph{radare2} understands the executable binary format, allowing it to process relocation and symbol information in particular. This allows us to extract symbols from the dynamic (\code{.dynsym}) as well as the static symbol table (\code{.symtab}) where present, and any strings referenced in the code. Our approach thus gains knowledge over functions of dynamic libraries used in the code. Finally, \emph{radare2} attempts to identify functions in code and generates corresponding control flow graphs.

\end{itemize}

Firstly, we strip the hexadecimal numbers from assembly instructions and replace them with the uni-gram $number$, to avoid overfitting that might be caused by unique hexadecimal numbers. Then, information provided by the two disassemblers is combined to obtain our disassembly feature set as follows: we tokenize the instruction traces of both disassemblers and extract token uni-grams, bi-grams, and tri-grams within a single line of assembly, and 6-grams, which span two consecutive lines of assembly. We cannot know exactly what each 6-gram corresponds to in assembly code but for most assembly instructions, a meaningful construct is longer than a line of assembly code. In addition, we extract single basic blocks of \emph{radare2}'s control flow graphs, as well as pairs of basic blocks connected by control flow.

%For both disassemblers, we subsequently tokenize their output, and create token uni-grams and bi-grams, which serve as our disassembly features.

%\textcolor{red}{Add diagram.}

%\textcolor{red}{Add diagram.}

\subsection{Feature extraction via decompilation}
\label{sec:decompilation}

\emph{Decompilers} are the second source of information that we consider for feature extraction in this work. In contrast to disassemblers, decompilers do not only uncover the program's machine code instructions, but additionally reconstruct higher level constructs in an attempt to translate an executable binary into equivalent source code. In particular, decompilers can reconstruct control structures such as different types of loops and branching constructs. We make use of these syntactical features of code as they have been shown to be valuable in the context of source code authorship attribution~\cite{caliskan2015anonymizing}. For decompilation, we employ the Hex-Rays decompiler \cite{hexrays}.
%, as well as a more simple open-source decompiler (Snowman~\cite{snowman}).

% \subsubsection{\textbf{Obtaining features from Hex-Rays}}

\begin{figure}[t]
\centering
 \includegraphics[height=35mm]{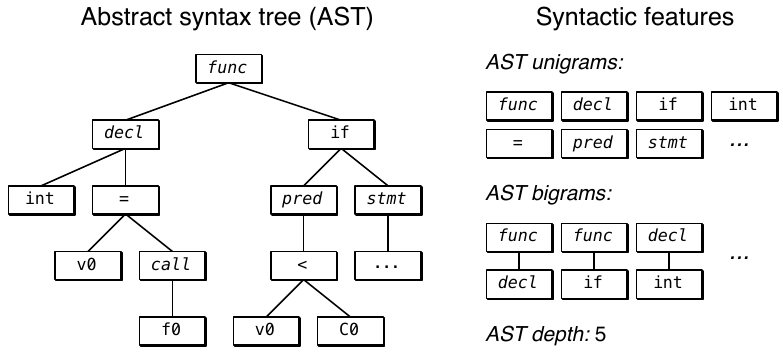}\\
 \vspace{6mm}
 \includegraphics[height=35mm]{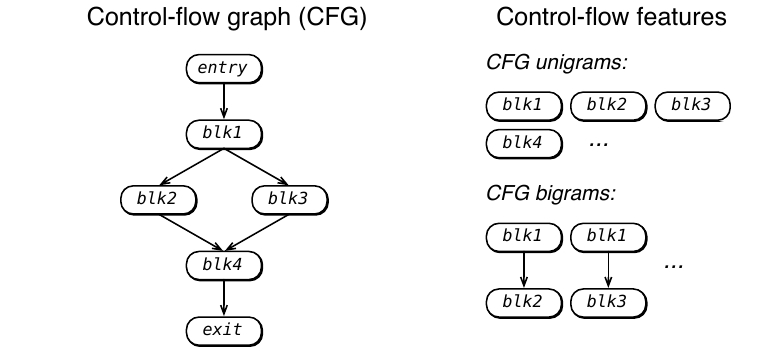}
 % \vspace{1cm}
\caption{\small{Feature extraction via decompilation and fuzzy parsing: C-like pseudo code produced by Hex-Rays is transformed into an abstract syntax tree and control-flow graph to obtain syntactic and control-flow features.}}
\label{fig:ast}
\end{figure}

Hex-Rays is a commercial state-of-the-art decompiler. It converts executable programs into a human readable C-like pseudo code to be read by human analysts. It is noteworthy that this code is typically significantly longer than the original source code. For example, decompiling an executable binary generated from $70$ lines of source code with Hex-Rays produces on average 900 lines of decompiled code. We extract two types of features from this pseudo code: lexical features, and syntactical features.
Lexical features are simply the word unigrams, which capture the integer types used in a program, names of library functions, and names of internal functions when symbol information is available. Syntactical features are obtained by passing the C-pseudo code to \emph{joern}, a fuzzy parser for C that is capable of producing fuzzy abstract syntax trees (ASTs) from Hex-Rays pseudo code output~\cite{YamGolArpRie14}. We derive syntactic features from the abstract syntax tree, which represent the grammatical structure of the program. Such features are (illustrated in Figure~\ref{fig:ast}) AST node unigrams, labeled AST edges, AST node term frequency inverse document frequency, and AST node average depth. Previous work on source code authorship attribution \cite{caliskan2015anonymizing, wisse2015scripting} shows that these features are highly effective in representing programming style.

% \subsubsection{\textbf{Obtaining CFG features from bjoern}}
% ADD
%\emph{Snowman} is an open source decompiler, which supports multiple different instruction sets and executable binary formats, including the Intel 32 bit instruction set and the ELF binary format considered in this work. In addition to decompiling code, \emph{Snowman} constructs control flow graphs. The nodes of the control flow graph are the \emph{basic blocks}, that is, sequences of statements that are known to always be executed in direct succession. These blocks provide a natural segmentation of a programs statements, which our method exploits. To this end, we normalize the code in basic blocks, and treat each basic block as a word. We then extract word uni-grams and bi-grams, that is, single basic blocks and sequences of two basic blocks, and apply the TF-IDF weighting scheme~\cite{} to obtain our final features.

\subsection{Dimensionality reduction}
\label{sec:dimreduc}
Feature extraction produces a large amount of features, resulting in sparse feature vectors with thousands of elements. However, not all features are equally informative to express a programmer's style. This makes it desirable to perform feature selection to obtain a compact representation of the data to reduce the computational burden during classification as well as the chances of overfitting. Moreover, sparse vectors may result in a large number of zero-valued attributes being selected during random forest's random subsampling of the attributes to select a best split. Reducing the dimensions of the feature set is important for avoiding overfitting. One example to overfitting would be a rare assembly instruction uniquely identifying an author. For these reasons, we use information gain criteria followed by correlation based feature selection to identify the most informative attributes that represent each author as a class. This reduces vector size and sparsity while increasing accuracy and model training speed. For example, we get 705,000 features from the 900 executable binary samples of 100 programmers. If we use all of these features in classification, the resulting de-anonymization accuracy is slightly above 30\% because the random forest might be randomly selecting features with values of zero in the sparse feature vectors. Once information gain criteria is applied, we get less than 2,000 features and the correct classification accuracy of 100 programmers increases from to 90\%. Then, we identify locally predictive features that are highly correlated with classes and have low intercorrelation. After this second dimensionality reduction method, we are left with 53 predictive features and no sparsity remains in the feature vectors. Extracting 53 features or training a machine learning model where each instance has 53 attributes is computationally efficient. Given such proper representation of instances, the correct classification accuracy of 100 programmers reaches 96\%.

We applied the first dimensionality reduction step using WEKA's information gain
attribute selection criterion~\cite{weka}, which evaluates the difference between the entropy
of the distribution of classes and the Shannon entropy of the conditional distribution
of classes given a particular feature~\cite{infogain}.

%\begin{equation}
%IG(A, M_{i}) = H(A) - H(A | M_{i})
%\end{equation}
%
%where $A$ is the class corresponding to an author, $H$ is Shannon entropy, and
%$M_{i}$ is the $i^{th}$ attribute of the data set. Intuitively, the
%Note that, as $H(A|M_{i})\leq H(A)$, information gain is always non-negative.
%While the use of information gain on a variable-per-variable basis implicitly
%assumes independence between the features with respect to their impact on the
%class label, this conservative approach to feature selection means that only
%those features that have demonstrable value in classification are included in
%our selected features.
%Information gain can be thought of as measuring the amount of information that
%the observation of the value of an attribute gives about the class label
%associated with the example. We retained only those features that individually had non-zero information gain.

The second dimensionality reduction step was based on correlation based feature selection, which generates a feature-class and feature-feature correlation matrix. The selection method then evaluates the worth of a subset of attributes by considering the individual predictive ability of each feature along with the degree of redundancy between them~\cite{hall1999correlation}. Feature selection is performed iteratively with greedy hillclimbing and backtracking ability by adding attributes that have the highest correlation with the class to the list of selected features.

%Subsets of features that are highly correlated with the class while having low intercorrelation are preferred. -- Identify locally predictive attributes. Iteratively adds attributes with the highest correlation with the class as long as there is not already an attribute in the subset that has a higher correlation with the attribute in question

%Searches the space of attribute subsets by greedy hillclimbing augmented with a backtracking facility. Setting the number of consecutive non-improving nodes allowed controls the level of backtracking done. Best first may start with the empty set of attributes and search forward, or start with the full set of attributes and search backward, or start at any point and search in both directions (by considering all possible single attribute additions and deletions at a given point).
%CFS subset eval generates a feature-class and feature-feature correlation matrix. Searches the space of attribute subsets by greedy hillclimbing augmented with a backtracking facility.

\subsection{Classification}
\label{sec:classification}

We used random forests as our classifier which are ensemble learners built from collections
of decision trees, where each tree is trained on a subsample of the
data obtained by random sampling with replacement. Random forests by nature are multi-class classifiers that avoid overfitting.
%randomly sampling \emph{N} training samples with replacement, where \emph{N} is the number of instances in the dataset.
%
 To reduce correlation between trees, features are also
 subsampled; commonly $(logM)+1$ features are selected at random
 (without replacement) out of $M$, and the best split on these
 $(logM)+1$ features is used to split the tree nodes.

 The number of selected features represents one of the few tuning parameters in
 random forests: increasing it increases the
 correlation between trees in the forest which can harm the accuracy of
 the overall ensemble, however increasing the number of features that
 can be chosen between at each split also increases the classification
 accuracy of each individual tree making them stronger classifiers with
 low error rates. The optimal range of number of features can be found
 using the out of bag error estimate, or the error estimate derived
 from those samples not selected for training on a given tree.

During classification, each test example is classified via each of the trained
decision trees by following the binary decisions made at each node until a leaf
is reached, and the results are aggregated. The most populous class is selected as the output of the forest for simple classification, or classifications can be ranked according to the number of trees that
`voted' for the label in question when performing relaxed attribution for \emph{top-n} classification.

We employed random forests with 500 trees, which empirically provided the best
tradeoff between accuracy and processing time. Examination of out of bag error
values across multiple fits suggested that $(logM)+1$ random features
(where $M$ denotes the total number of features) at each split of the
decision trees was in fact optimal in all of the experiments listed in Section
\ref{sec:experiments}, and was used throughout. Node splits were selected based on
the information gain criteria, and all trees were
grown to the largest extent possible, without pruning.

The data was analyzed via \emph{k}-fold cross-validation, where the data was
split into training and test sets stratified by author (ensuring that the
number of code samples per author in the training and test sets was identical
across authors). The parameter \emph{k} varies according to datasets and is equal to the number of instances present from each author. The cross-validation procedure was repeated 10 times, each with a different random seed, and average results across all iterations are reported, ensuring that results are not biased by improbably easy or difficult to classify subsets.

We report our classification results in terms of kappa statistics, which is roughly equivalent to accuracy but subtracts the random chance of correct classification from the final accuracy. As programmer de-anonymization is a multi-class classification problem, an evaluation based on accuracy, or the true positive rate, represents the correct classification rate in the most meaningful way.

%% file: 6_experiments.tex
\section{Google Code Jam experiments}
\label{sec:experiments}

In this section, we go over the details of the various experiments we performed to address the research question formulated in Section~\ref{sec:problem}.

\input{4_data}

\subsection{\textbf{53 features represent programmer style.}}

We are interested in identifying features that represent coding style preserved in executable binaries. With the current approach, we extract 705,000 representations of code properties of 100 authors, but only a subset of these are the result of individual programming style. We are able to capture the features that represent each author's programming style that is preserved in executable binaries by applying information gain criteria to these 705,000 features. After applying information gain to effectively represent coding style, we reduce the feature set to contain approximately 1,600 features from all feature types. Furthermore, correlation based feature selection during cross validation eliminates features that have low class correlation and high intercorrelation and preserves 53 of the highly distinguishing features which can be seen in Table~\ref{tab:ig} along with their authorial style representation power. 

Considering the fact that we are reaching such high accuracies on de-anonymizing 100 programmers with 900 executable binary samples (discussed below), these features are providing strong representation of style that survives compilation. The compact set of identifying stylistic features contain features of all types, namely disassembly, CFG, and syntactical decompiled code properties. To examine the potential for overfitting, we consider the ability of this feature set to generalize to a different set of programmers (see Section~\ref{subsection:nooverfitting}), and show that it does so, further supporting our belief that these features effectively capture coding style. Features that are highly predictive of authorial fingerprints include file and stream operations along with the formats and initializations of variables from the domain of ASTs whereas arithmetic, logic, and stack operations are the most distinguishing ones among the assembly instructions.

\begin{table}[!htbp]
\footnotesize
\begin{center}
\begin{tabular}{| p{2.5cm} |  p{1.3cm} | P{1.3cm} | P{2.2cm} |} \hline
\textbf{Feature} & \textbf{Source}&  \textbf{Number of Possible Features}  &\textbf{Selected Features \newline / \newline Information Gain}  \\ \hline

 {Word unigrams} & decompiled\newline code$^{*}$ &  29,278 & 6/5.75   \\\hline
  {AST\newline node TF\dag } & decompiled\newline code$^{*}$ & 5,278 &  3/1.85  \\\hline
  {Labeled AST\newline edge TF\dag} &  decompiled\newline code$^{*}$ & 26,783  & 0/0 \\\hline
 {AST\newline node TFIDF\ddag} & decompiled\newline code$^{*}$ & 5,278 & 1/0.75  \\\hline
 {AST node\newline average depth} & decompiled\newline code$^{*}$ &5,278  & 0/0 \\\hline
 {C++ keywords} & decompiled\newline code$^{*}$ & 73 & 0/0 \\\hline
 radare2\newline disassembly unigrams & radare\newline disassembly & 21,206 & 3/1.61  \\\hline
 radare2 disassembly bigrams &radare\newline disassembly & 39,506 &1/0.62 \\\hline
  radare2 disassembly trigrams &radare\newline disassembly &  112,913& 0/0 \\\hline
      radare2 disassembly 6-grams &ndisasm\newline disassembly & 260,265 & 0/0 \\\hline

radare2 CFG\newline node unigrams & radare\newline disassembly & 5,297  & 3/1.98 \\\hline
radare2\newline CFG edges &radare\newline disassembly & 10,246 & 1/0.63 \\\hline

  ndisasm disassembly unigrams & ndisasm\newline disassembly & 5,383  & 2/1.79   \\\hline
  ndisasm disassembly bigrams &ndisasm\newline disassembly & 14,305  & 5/2.95 \\\hline
    ndisasm disassembly trigrams &ndisasm\newline disassembly & 5,237 & 4/1.44\\\hline
    ndisasm\newline disassembly 6-grams &ndisasm\newline disassembly & 159,142 & 24/16.08\\\hline

\multicolumn{2}{|c|}{{Total}}   & 705,468&  53/35  \\\hline
\multicolumn{2}{|c|}{\textit{$^{*} $hex-rays decompiled code}}  & \multicolumn{2}{|c|}{\textit{TF\dag = term frequency}}   \\\hline
\multicolumn{4}{|c|}{\textit{TFIDF\ddag = term frequency inverse document frequency}}   \\ \hline
\end{tabular}
\end{center}
\captionsetup{justification=centering,margin=0cm}
\caption{ \label{tab:ig} Programming Style Features and\newline Selected Features in Executable Binaries}
\vspace{-1.0cm}

\end{table}

\subsection{\textbf{We can de-anonymize programmers from their executable binaries.}}
\label{sec:mainexperiment}

This is the main experiment that demonstrates how de-anonymizing programmers from their executable binaries is possible. After preprocessing the dataset to generate the executable binaries without optimization, we further process the executable binaries to obtain the disassembly, control flow graphs, and decompiled source code. We then extract all the possible features detailed in Section~\ref{sec:approach}. We take a set of 100 programmers who all have 9 executable binary samples. With 9-fold-cross-validation, the random forest correctly classifies 900 test instances with 95\% accuracy, which is significantly higher than the accuracies reached in previous work.

There is an emphasis on the number of folds used in these experiments because each fold corresponds to the implementation of the same algorithmic function by all the programmers in the GCJ dataset (e.g. all samples in fold 1 may be attempts by the various authors to solve a list sorting problem). Since we know that each fold corresponds to the same Code Jam problem, by using stratified cross validation without randomization and preserving order, we ensure that all training and test samples contain the same algorithmic functions implemented by all of the programmers. The classifier uses the excluded fold in the testing phase, which contains executable binary samples that were generated from an algorithmic function that was \emph{not} previously observed in the training set for that classifier. Consequently, the only distinction between the test instances is the coding style of the programmer, without the potentially confounding effect of identifying an algorithmic function.

%\begin{table}[!htbp]
%\begin{center}
%\begin{tabular}{| p{1.5cm} | p{1.5cm} | p{2cm} | p{1.5cm}|} \hline
%\textbf{Number of Programmers} & \textbf{Number of Training Samples} & \textbf{Cross Validation} &\textbf{Accuracy} \\ \hline
% {20} & 5 & 5-fcv* &{95.0\%} \\\hline
% {20} & 13 & 13-fcv* &{96.0\%} \\\hline
% {20} & 8 & 8-fcv* & {96.0\%} \\\hline
% {100} & {8} & 8-fcv* & {78.3\%} \\\hline
%\multicolumn{4}{|l|} {\textit{*k-fcv refers to k-fold cross validation}}\\\hline
%\end{tabular}
%\end{center}
%\caption{ \label{tab:prog} Programmer De-anonymization}
%%\vspace{-0.75cm}
%\end{table}

\subsection{\textbf{Even a single training sample per programmer is sufficient for de-anonymization.}}
\label{subsec:data}

\begin{figure}
\begin{center}
\begin{tikzpicture}[scale=0.65]
	\begin{axis}[
	height=6cm,
	width=10cm,
		xlabel=Number of Training Samples Per Author,
		 ylabel=Correct Classification Accuracy,
		 xtick={1, 2, 3,4,5,6,7,8},
		 ymin =60, ymax=100,
		 xmin=0, xmax=9,
		 minor tick num=5,
		 grid=both,
 grid style={line width=.1pt, draw=gray!10},
 major grid style={line width=.2pt,draw=gray!50},
 ]
	\addplot[color=red,mark=x] coordinates {
	 (1 ,65 )
 ( 2 ,85)
 ( 3 ,91)
 ( 4 ,91)
 ( 5 , 94)
 ( 6 ,94)
 (7 , 95)
  (8 , 96)

	};
		\node at ( axis cs:0.7,68){$65\%$};
		\node at ( axis cs:1.9,88){$85\%$};
		\node at ( axis cs:3,93){$91\%$};
		\node at ( axis cs:4,94){$91\%$};
		\node at ( axis cs:5,96){$94\%$};
		\node at ( axis cs:6,96){$94\%$};
		\node at ( axis cs:7,98){$95\%$};
		\node at ( axis cs:8,98){$96\%$};

	\end{axis}
\end{tikzpicture}
\captionsetup{justification=centering,margin=0cm}

	\caption{ \label{fig:reqdata} {Amount of Training Data Required for De-anonymizing 100 Programmers}}
\end{center}
\vspace{-1.0cm}
\end{figure}
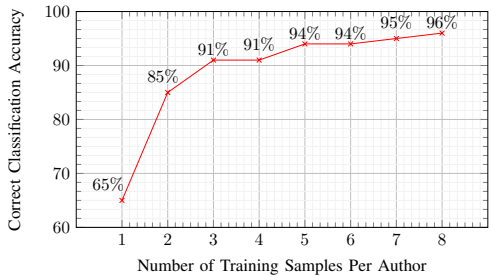

A drawback of supervised machine learning methods, which we employ, is that they require labeled examples to build a model. The ability of the model to accurately generalize is often strongly linked to the amount of data provided to it during the training phase, particularly for complex models. In domains such as executable binary authorship attribution, where samples may be rare and obtaining ``ground truth'' for labeling training samples may be costly or laborious, this can pose a significant challenge to the usefulness of the method.

We therefore devised an experiment to determine how much training data is required to reach a stable classification accuracy, as well as to explore the accuracy of our method with severely limited training data. As programmers produce a limited number of code samples per round of the GCJ competition, and programmers are eliminated in each successive round, the GCJ dataset has an upper bound in the number of code samples per author as well as a limited number of authors with a large number of samples. Accordingly, we identified a set of 100 programmers that had exactly 9 program samples each, and examined the ability of our method to correctly classify each author out of the candidate set of 100 authors when training on between 1 and 8 files per author.

As shown in Figure~\ref{fig:reqdata}, the classifier is capable of correctly identifying the author of a code sample from a potential field of 100 with 65\% accuracy on the basis of a single training sample. The classifier also reaches a point of dramatically diminishing returns with as few as three training samples, and obtains a stable accuracy by training on 6 samples. Given the complexity of the task, this combination of high accuracy with extremely low requirement on training data is remarkable, and suggests the robustness of our features and method. It should be noted, however that this set of programmers with a large number of files corresponds to more skilled programmers, as they were able to remain in the competition for a longer period of time and thus produce this large number of samples.

\subsection{\textbf{Relaxed Classification: In difficult scenarios, the classification task can be narrowed down to a small suspect set.}}
\label{sec:relaxed}

\begin{figure}
\begin{center}
\begin{tikzpicture}[scale=0.6]
	\begin{axis}[legend style={at={(0.5,-0.22)},anchor=north},
	height=6cm,
	width=10cm,
		xlabel=n,
		 ylabel=Correct Classification Accuracy,
		 xtick={1, 2, 3,4,5,6,7,8,9,10},
		 ymin =0, ymax=100,
		 xmin=0, xmax=11,
		 minor tick num=5,
		 grid=both,
 grid style={line width=.1pt, draw=gray!10},
 major grid style={line width=.2pt,draw=gray!50},
 ]

	\addplot[color=red,mark=x] coordinates {
	 (1 ,96 )
 ( 2 ,98)
 ( 3 ,99)
 ( 4 ,99)
 ( 5 , 99)
 (6, 99)
  (7, 99)
 (8, 99)
 (9, 99)
  (10, 99)
  	};
	  \addlegendentry{\textit{100 programmers 8 training samples}}

	  	\addplot[color=green,mark=x] coordinates {

  	 (1 ,66 )
 ( 2 ,76)
 ( 3 ,78)
 ( 4 ,83)
 ( 5 , 83)
 (6, 86)
  (7, 87)
 (8, 87)
 (9, 87)
  (10, 88)
  	};
	  \addlegendentry{\textit{100 programmers 1 training sample}}

		  	\addplot[color=purple,mark=x] coordinates {

  	 (1 ,83 )
 ( 2 ,88)
 ( 3 ,89)
 ( 4 ,90)
 ( 5 , 91)
 (6, 92)
  (7, 92)
 (8, 93)
 (9, 93)
  (10, 93)
  	};

	  \addlegendentry{\textit{600 programmers 8 training samples}}
	  		\addplot[color=blue,mark=x] coordinates {
		  	 (1 ,14 )
 ( 2 ,21)
 ( 3 ,25)
 ( 4 ,30)
 ( 5 ,33)
 (6, 36)
  (7, 39)
 (8, 41)
 (9, 43)
  (10,45 )
	};
	
		  \addlegendentry{\textit{600 programmers 1 training sample}}

%		\node at ( axis cs:0.7,68){$65\%$};
%		\node at ( axis cs:1.9,88){$85\%$};
%		\node at ( axis cs:3,93){$91\%$};
%		\node at ( axis cs:4,94){$91\%$};
%		\node at ( axis cs:5,96){$94\%$};
%
	\end{axis}

\end{tikzpicture}
\captionsetup{justification=centering,margin=0cm}
	\caption{ \label{fig:relaxed100} {Reducing Suspect Set Size: \newline Top-n Relaxed Classification}}
\end{center}
\vspace{-1.0cm}
\end{figure}
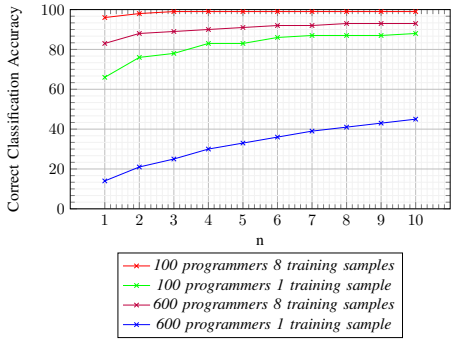

In Section~\ref{sec:dataset}, the previously unseen anonymous executable binary sample is classified such that it belongs to the most likely author's class. In cases where we have too many classes or the classification accuracy is lower than expected, we can relax the classification to {\em top--n} classification. In {\em top--n} relaxed classification, if the test instance belongs to one of the most likely $n$ classes, the classification is considered correct. This can be useful in cases when an analyst or adversary is interested in finding a suspect set of $n$ authors, instead of a direct {\em top--1} classification. Being able to scale down an authorship investigation for an executable binary sample of interest to a reasonable sized set of suspect authors among hundreds of authors greatly reduces the manual effort required by an analyst or adversary. Once the suspect set size is reduced, the analyst or adversary could adhere to content based dynamic approaches and reverse engineering to identify the author of the executable binary sample. Figure~\ref{fig:relaxed100} shows how correct classification accuracies approach 100\% as the classification is relaxed to top-10.

%\begin{figure}[htbp]
%\centering
% \includegraphics[width=0.49\textwidth]{figs/relaxed20programmers.png}
%\caption{De-anonymizing 20 Programmers}
%\label{fig:dean20}
%\end{figure}
%\begin{figure}[htbp]
%\begin{center}
% \includegraphics[width=0.50\textwidth]{figs/relaxed100programmersOptimized.png}
%\caption{De-anonymizing 100 Programmers}
%\label{fig:dean100}
%
%\end{center}
%\end{figure}

It is important to note from Figure~\ref{fig:reqdata} that, by using only a single training sample in a 100-class classification task, the machine learning model can correctly classify new samples with 75.0\% accuracy. This is of particular interest to an analyst or adversary who does not have a large amount of labeled samples in her suspect set. Figure~\ref{fig:reqdata} shows that an analyst or adversary can narrow down the suspect set size from 100 or 600 to a significantly smaller set.

%\begin{figure}[htbp]
%\begin{center}
% \includegraphics[width=0.49\textwidth]{figs/oneTrainingSample.png}
%\captionsetup{justification=centering,margin=0cm}
%\caption{De-anonymizing 20 Programmers with One Training Sample}
%\label{fig:oneTrain}
%\end{center}
%\end{figure}

%\subsection{Feature Validation}
\subsection{\textbf{The feature set selected via dimensionality reduction works and is validated across different sets of programmers.}}
\label{subsection:nooverfitting}

%\aylin{examples of 53 features}

In our earlier experiments, we trained the classifier on the same set of executable binaries that we used during feature selection. The high number of starting features from which we select our final feature set via dimensionality reduction does raise the potential concern of overfitting. To examine this, we applied this final feature set to a different set of programmers and executable binaries. If we reach accuracies similar to what we got earlier, we can conclude that these selected features do generalize to other programmers and problems, and therefore are not overfitting to the 100 programmers they were generated from. This also suggests that the final set of features in general capture programmer style.

Recall that analyzing 900 executable binary samples of the 100 programmers resulted in about 705,000 features, and after dimensionality reduction, we are left with 53 important features. We picked a different (non-overlapping) set of 100 programmers and performed another de-anonymization experiment in which the feature selection step was omitted, using instead the information gain and correlation based features obtained from the original experiment. This resulted in very similar accuracies: we de-anonymized programmers in the validation set with 96\% accuracy by using features selected via the main development set, compared to the 95\% de-anonymization accuracy we achieve on the programmers of the main development set. The ability of the final reduced set of 53 features to generalize beyond the dataset which guided their selection strongly supports the assertion that these features obtained from the main set of 100 programmers are not overfitting, and they actually represent coding style in executable binaries, and can be used across different datasets.

\subsection{\textbf{Large Scale De-anonymization: We can de-anonymize 600 programmers from their executable binaries.}}
\begin{figure}
\begin{center}
\begin{tikzpicture}[scale=0.6]
	\begin{axis}[
	height=6cm,
	width=10cm,
		xlabel=Number of Authors,
		 ylabel=Correct Classification Accuracy,
		 xtick={20, 100, 200,300,400,500,600},
		 ymin =0, ymax=100,
		 xmin=0, xmax=600,
		 minor tick num=5,
		 grid=both,
 grid style={line width=.1pt, draw=gray!10},
 major grid style={line width=.2pt,draw=gray!50},
 ]
	\addplot[color=red,mark=x] coordinates {
	 (600 ,82.6 )
 ( 500 ,83.2)
 ( 400 ,85.2)
 ( 300 ,89.0)
 ( 200 , 91.7)
 ( 100 ,95.7)
 (20 , 99.6)
	};
		\node at ( axis cs:30,95){$99\%$};
		\node at ( axis cs:100,92){$96\%$};
		\node at ( axis cs:200,88){$92\%$};
		\node at ( axis cs:300,85){$89\%$};
		\node at ( axis cs:400,80){$85\%$};
		\node at ( axis cs:500,78){$83\%$};
		\node at ( axis cs:570,78){$83\%$};

	\end{axis}
\end{tikzpicture}
\captionsetup{justification=centering,margin=0cm}
	\caption{ \label{fig:large} { Large Scale Programmer De-anonymization}}
\end{center}
\vspace{-1.0cm}
\end{figure}
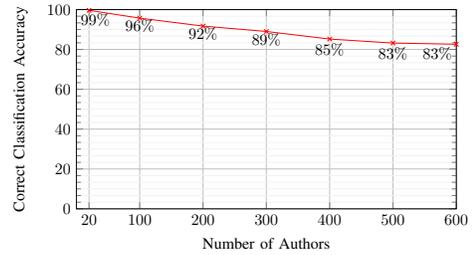

We would like to see how well our method scales up to 600 users. An analyst with a large set of labeled samples might be interested in performing large scale de-anonymization. For this experiment, we use 600 contestants from GCJ with 9 files. We only extract the reduced set of features from the 600 users. This decreases the amount of time required for feature extraction. On the other hand, this experiment shows how effectively overall programming style is represented after dimensionality reduction. The results of large scale programmer de-anonymization in Figure~\ref{fig:large}, show that our method can scale to larger datasets with the reduced set of features with a surprisingly small drop on accuracy.
%\begin{figure}[htbp]
%\begin{center}
% \includegraphics[width=0.48\textwidth]{figs/largeScale_new.png}
%\caption{Large Scale Programmer De-anonymization}
%\label{fig:large}
%
%\end{center}
%\end{figure}
%
%\begin{quote}\small
%\begin{bchart}[min=0,step=20,max=100, unit=\%,scale=0.8]
%
% \bcbar[text=600 authors, color = green!13!gray]{82.6 } 
% \bcbar[text= 500 authors, color = green!16!gray ]{83.2}
% \bcbar[text= 400 authors, color = green!26!gray ]{85.2}
% \bcbar[text= 300 authors , color = green!45!gray]{89.0}
% \bcbar[text= 200 authors, color = green!58!gray]{91.7}
% \bcbar[text= 100 authors, color = green!78!gray]{95.7}
% \bcbar[text={\hspace{0.07cm} 20 authors}, color = green!98!gray]{99.6}
% \bcxlabel{Correct Classification Accuracy}
%
%
%\end{bchart}
%\end{quote}

\subsection{\textbf{We advance the state of executable binary authorship attribution.}}

Rosenblum et al. presented the largest scale evaluation of executable binary authorship attribution on 191 programmers each with at least 8 training samples \cite{rosenblum2011wrote}. We compare our results with Rosenblum et al.'s in Table~\ref{tab:comp} to show how we advance the state of the art both in accuracy and on larger datasets. Rosenblum et al. use 1,900 coding style features to represent coding style whereas we use 53 features, which might suggest that our features are more powerful in representing coding style that is preserved in executable binaries. On the other hand, we use less training samples as opposed to Rosenblum et al., which makes our experiments more challenging from a machine learning standpoint. Our accuracy in authorship attribution is significantly higher than Rosenblum et al.'s, even when we use an SVM as our classifier, showing that our different approach is more powerful and robust for de-anonymizing programmers. Rosenblum et al. suggest a linear SVM is the appropriate classifier for de-anonymizing programmers but we show that our different set of techniques and choice of random forests is leading to superior and larger scale de-anonymization.

\begin{table}[!htbp]
\small
\begin{center}
\begin{tabular}{| p{2.1cm} | p{1.2cm} | p{1.73cm} | P{0.8cm}| P{0.8cm}|} \hline
\scriptsize{Related Work} &
\scriptsize{Number of \newline Programmers} &
\scriptsize{Number of \newline Training Samples} &
\scriptsize{Accuracy} &\scriptsize{Classifier}\\ \hline

%Rosenblum \cite{rosenblum2011wrote}& \textbf{20}& 8-16 & {77\%} & SVM\\\hline
% This work & \textbf{100} & 8 & {96\%} & RF \\\hline
%\hline
%\hline

Rosenblum \cite{rosenblum2011wrote}& 20& \textbf{8-16} & {77\%} & SVM\\\hline
This work & 20 & \textbf{8} &{90\%} & SVM\\\hline
This work & 20 & \textbf{8} &{99\%} & RF\\\hline
\hline
\hline

Rosenblum \cite{rosenblum2011wrote}& {100} & 8-16&\textbf{61\%} & SVM\\\hline
This work& {100} & 8 & \textbf{84\%} & SVM\\\hline
This work& {100} & 8 & \textbf{96\%} & RF\\\hline
\hline
\hline

Rosenblum \cite{rosenblum2011wrote}& 191 & 8-16&\textbf{51\%} & SVM\\\hline
This work & 191 & 8 & \textbf{81\%}& SVM\\\hline
This work & 191 & 8 & \textbf{92\%}& RF\\\hline
This work & \textbf{600} & 8 & \textbf{71\%}& SVM\\\hline
This work & \textbf{600} & 8 & \textbf{83\%}& RF\\\hline

\end{tabular}
\end{center}

\captionsetup{justification=centering,margin=0cm}

\caption{ \label{tab:comp} Comparison to Previous Results}
\vspace{-0.5cm}
\end{table}

\subsection{\textbf{Programmer style is preserved in executable binaries.}}

\label{subsec:reconstruction}
We show throughout the results that it is possible to de-anonymize programmers from their executable binaries with a high accuracy. To quantify how stylistic features are preserved in executable binaries, we calculated the correlation of stylistic source code features and decompiled code features. We used the stylistic source code features from previous work on de-anonymizing programmers from their source code \cite{caliskan2015anonymizing}. We took the most important 150 features in coding style that consist of AST node average depth, AST node TFIDF, and the frequencies of AST nodes, AST node bigrams, word unigrams, and C++ keywords. For each executable binary sample, we have the corresponding source code sample. We extract 150 information gain features from the original source code. We extract decompiled source code features from the decompiled executable binaries. For each executable binary instance, we set one corresponding information gain feature as the class to predict and then we calculate the correlation between the decompiled executable binary features and the class value. A random forest classifier with 500 trees predicts the class value of each instance, and then Pearson's correlation coefficient is calculated between the predicted and original values. The correlation has a mean of 0.32 and ranges from -0.12 to 0.69 for the most important 150 features.

To see how well we can reconstruct the original source code features from decompiled executable binary features, we reconstructed the 900 instances with 150 features that represent the highest information gain features by predicting the original features from decompiled code features. We calculated the cosine similarity between the original 900 instances and the reconstructed instances after normalizing the features to unit distance. The cosine similarity for these instances is in Figure~\ref{fig:optim}, where a cosine similarity of 1 means the two feature vectors are identical. The high values (average of 0.81) in cosine similarity suggest that the reconstructed features are similar to the original features. When we calculate the cosine similarity between the feature vectors of the original source code and the corresponding decompiled code's feature vectors (\emph{no predictions}), the average cosine similarity is 0.35. In summary, reconstructed features are much more similar to original code than the raw features extracted from decompiled code. 5\% of the reconstructed features have less than 60\% similarity based on the cosine similarity between original and decompiled source code features. At the same time, the de-anonymization accuracy of 900 executable binaries is 95\% by using source code, assembly, CFG, and AST features. This might indicate that some operations or code sequences cannot be preserved after compilation followed by decompilation, due to the nature of transformations during each process.

\input{reconstructedFeatures}

\section{Real-World Scenarios}
\label{sec:compExp}

%\subsection{textbf{Effect of Compiler Optimization on Programmer De-anonymization}
\subsection{\textbf{Programmers of optimized executable binaries can be de-anonymized.}}

In Section~\ref{sec:experiments}, we discussed how we evaluated our approach on a controlled and clean real-world dataset. Section~\ref{sec:experiments} shows how we advance over previous methods that were all evaluated with clean datasets such as GCJ or homework assignments. In this section, we investigate a complicated dataset which has been optimized during compilation, where the executable binary samples have been normalized further during compilation.

Compiling with optimization tries to minimize or maximize some attributes of an executable program. The goal of optimization is to minimize execution time or the amount of memory a program occupies. The compiler applies optimizing transformations which are algorithms that transform a program to a semantically equivalent program that uses fewer resources.

GCC has predefined optimization levels that turn on sets of optimization flags. Compilation with optimization level-1, tries to reduce code size and execution time, takes more time and much more memory for large functions than compilation with no optimizations. Compilation with optimization level-2 optimizes more than level-1, uses all level-1 optimization flags and more. Level-2 optimization performs all optimizations that do not involve a space-speed tradeoff. Level-2 optimization increases compilation time and performance of the generated code when compared to level-1 optimization. Level-3 optimization yet optimizes more than both level-1 and level-2.

So far, we have shown that programming style features survive compilation without any optimizations. As compilation with optimizations transforms code further, we investigate how much programming style is preserved in executable binaries that have gone through compilation with optimization. Our results summarized in Table~\ref{tab:opt} show that programming style is preserved to a great extent even in the most aggressive level-3 optimization. This shows that programmers of optimized executable binaries can be de-anonymized and optimization is not a highly effective code anonymization method.

\begin{table}[!htbp]
\footnotesize
\begin{center}
\begin{tabular}{| p{2.1cm} | p{1.6cm} | p{1.98cm} | p{1.3cm}|} \hline
\textbf{\small{Number of Programmers}} & \textbf{\small{Number of Training Samples}} &\textbf{\small{Compiler Optimization Level}} & \textbf{\small{Accuracy}} \\ \hline

 {100} & {8}& None & {96\%} \\\hline
 {100} & {8}& 1 & {93\%} \\\hline
 {100} & {8}& 2 & {89\%} \\\hline
 {100} & {8}& 3 & {89\%} \\\hline

 \hline

\end{tabular}
\end{center}
\captionsetup{justification=centering,margin=0cm}
\caption{ \label{tab:opt} Programmer De-anonymization with Compiler Optimization}
\vspace{-0.5cm}

\end{table}

\subsection{\textbf{Removing symbol information does not anonymize executable binaries.}}

To investigate the relevance of symbol information for classification
accuracy, we repeat our experiments with $100$ authors presented in
the previous section on \emph{fully stripped executable binaries}, that is,
executable binaries where symbol information is missing completely. We obtain
these executable binaries using the standard utility \emph{GNU strip} on each
executable binary sample prior to analysis. Upon removal of symbol information, without any optimizations, we
notice a decrease in classification accuracy by 24\%, showing that stripping symbol information from executable binaries is not effective enough to anonymize an executable binary sample.

\subsection{\textbf{We can de-anonymize programmers from obfuscated binaries.}}
\label{sec:obfuscationExp}

We are furthermore interested in finding out whether our method is
capable of dealing with simple binary obfuscation techniques as
implemented by tools such as
Obfuscator-LLVM~\cite{ieeespro2015-JunodRWM}. These obfuscators substitute instructions by other semantically equivalent
instructions, they introduce bogus control flow, and can even
completely flatten control flow graphs.

For this experiment, we consider a set of 100 programmers from the GCJ
data set, who all have 9 executable binary samples. This is the same
data set as considered in our main experiment (see
Section~\ref{sec:mainexperiment}), however, we now apply all three
obfuscation techniques implemented by Obfuscator-LLVM to the samples
prior to learning and classification.

We proceed to train a classifier on obfuscated samples. This approach
is feasible in practice as an analyst who has only non-obfuscated
samples available can easily obfuscate them to obtain the necessary
obfuscated samples for classifier training. Using the same features as in Section~\ref{sec:mainexperiment}, we obtain an
accuracy of 88\% in correctly classifying authors.

\subsection{\textbf{De-anonymization \emph{in the Wild}}}
\label{sec:gitExp}

To better assess the applicability of our programmer de-anonymization
approach \emph{in the wild}, we extend our experiments to code
collected from real open-source programs as opposed to solutions for
programming competitions. To this end, we automatically collected
source files from the popular open-source collaboration platform
GitHub \cite{git}. Starting from a seed set of popular repositories, we traversed the platform to obtain C/C++ repositories that meet the
following criteria. Only one author has committed to the repository. The repository is popular as indicated by the presence of at
 least 5 \emph{stars}, a measure of popularity for repositories on
 GitHub. Moreover, it is sufficiently large, containing a total of 200
 lines at least. The repository is not a fork of another repository, nor is it
 named `linux', `kernel', `osx', `gcc', `llvm', `next', as these
 repositories are typically copies of the so-named projects.

We cloned 439 repositories from 161 authors meeting these criteria and collect only C/C++ files for which the main author has contributed at least 5
commits and the commit messages do not contain the word 'signed-off',
a message that typically indicates that the code is written by another
person. An author and her files are included in the dataset only if she has written at least 10 different
files. In the final step, we manually verified ground truth on authorship for the selected files to make
sure that they do not show any clear signs of code reuse from other projects. The resulting dataset had 2 to 344 files and 2 to 8 repositories from each author, with a total of 3,438 files.
%

%The remaining files are indexed and further filtered using the identified authors.
% fabs: I've commented this out. It was very relevant for the initial
% github experiments because we were looking at entire binaries, but
% now that we're looking at object files, I think it no longer is. In
% particular because of the manual cleaning performed.

 We developed our method and evaluated it on the GCJ dataset, but collecting code from open source projects is another option for constructing a dataset. Open source projects do not guarantee ground truth on authorship. The feature vectors might capture topics of the project instead of programming style. As a result, open source code does not constitute the ideal data for authorship analysis; however, it allows us to better assess the applicability of programmer de-anonymization in the wild. We therefore present results from a dataset collected from the hosting platform GitHub, which we obtain by spidering the platform to collect C and C++ repositories.
%

%\begin{table}[!htbp]
%\small
%\begin{center}
%\begin{tabular}{ | p{3.3cm} | p{2.2cm}| }
%
%\hline
%{\textbf{Type}} & \textbf{Amount} \\ \hline
%Authors & 161\\ \hline
%Repositories & 439 \\ \hline
%Files & 3,438 \\ \hline
%Repositories / Author & 2 -- 8 \\ \hline
%Files / Author & 2 -- 344 \\ \hline
%\end{tabular}
%\end{center}
%
%\caption{\label{tab:git1} Single Authored GitHub Repositories}
%\end{table}

% {\textbf{We can de-anonymize GitHub programmers.}}

We subsequently compile the collected projects to obtain object files
for each of the selected source files. We perform our experiment on
object files as opposed to entire binaries, since the object files are the binary
representations of the source files that clearly belong to the specified authors.

For different reasons, compiling code may not be possible for a
project, e.g., the code may not be in a compilable state, it may not
be compilable for our target platform (32 bit Intel, Linux), or the files to setup a working build environment can no longer be
obtained. Despite these difficulties, we are able to generate 1,075
object files from 90 different authors, where the number of object
files per author ranges from 2 to 24, with most authors having at least 9
samples. We used 50 of these authors that have 6 to 15 files to perform a machine learning experiment with more balanced class sizes.

We extract the information gain features that were selected from GCJ data from
this GitHub dataset. GitHub datasets are noisy for two reasons since the executable binaries used in de-anonymization might contain properties from third party libraries and code. For these two reasons, it is more difficult to attribute authorship to anonymous executable binary samples from GitHub, but nevertheless we reach 65\% accuracy in correctly classifying these programmers' executable binaries. Another difficulty in this particular dataset is that there is not much training data to train an accurate random forest classifier that models each programmer. For example, we can de-anonymize the two programmers with the most samples, one with 11 samples and one with 7, with 100\% accuracy.

Being able to de-anonymize programmers in the wild by using a small number of features obtained from our clean development dataset is a promising step towards attacking more challenging real-world de-anonymization problems. %to be precise, I use 58 features after correlation based feature selection, but using all of IG features leads to slightly higher accuracy

\subsection{\textbf{Have I seen this programmer before?}}
\label{sec:openworld}

While attempting to de-anonymize programmers in real-world settings, we cannot be certain that we have formerly encountered code samples from the programmers in the test set. As a mechanism to check whether an anonymous test file belongs to one of the candidate programmers in the training set, we extend our method to an open world setting by incorporating classification confidence thresholds. In random forests, the class probability or classification confidence $P(B_{i})$ that executable binary $B$ is of class $i$ is calculated by taking the percentage of trees in the random forest that voted for class $i$ during classification.
\begin{equation} \label{conf}
P(B_{i}) = \frac{ \sum_j {V_j(i)}}{ |T|_{f} }
\end{equation}

There are multiple ways to assess classifier confidence and we devise a method that calculates the classification confidence by using classification margins. In this setting, the classification margin of a single instance is the difference between the highest and second highest $P(B_{i})$. The first step towards attacking an open world classification task is identifying the confidence threshold of the classifier for classification verification. As long as we determine a confidence threshold based on training data, we can calculate the probability that an instance belongs to one of the programmers in the training set and accordingly accept or reject the classification.

We performed 900 classifications in a 100-class problem to determine the confidence threshold based on the training data. The accuracy was 95\%. There were 40 misclassifications with an average classification confidence of 0.49. We took another set of 100 programmers with 900 samples. We classify these 900 samples with the closed world classifier that was trained in the first step on samples from a disjoint set of programmers. All of the 900 samples are attributed to a programmer in the closed world classifier with a mean classification confidence of 0.40. We can pick a verification threshold and reject all classifications with confidence below the selected threshold. Accordingly all the rejected open world samples and misclassifications become true negatives, and the rejected correct classifications end up as false negatives. Open world samples and misclassifications above the threshold are false positives and the correct classifications are true positives. Based on this, we generate an accuracy, pecision, and recall graph with varying confidence threshold values in Figure~\ref{fig:open}. This figure shows that the optimal rejection threshold to guarantee 90\% accuracy on 1,800 samples and 100 classes is around confidence 0.72. Other confidence thresholds can be picked based on precision and recall trade-offs. These results are encouraging for extending our programmer de-anonymization method to open world settings where an analyst deals with many uncertainties under varying fault tolerance levels.

\input{openWorld}

The experiments in this section can be used in software forensics to find out the programmer of a piece of malware. In software forensics, the analyst does not know if source code belongs to one of the programmers in the candidate set of programmers. In such cases, we can classify the anonymous source code, and if the majority number of votes of trees in the random forest is below a certain threshold, we can reject the classification considering the possibility that it might not belong to any of the classes in the training data. By doing so, we can scale our approach to an open world scenario, where we might not have encountered the suspect before. As long as we determine a confidence threshold based on training data \cite{stolerman2014classify}, we can calculate the probability that an instance belongs to one of the programmers in the set and accordingly accept or reject the classification.
We performed 270 classifications in a 30-class problem using all the features to determine the confidence threshold based on the training data. The accuracy was 96.67\%. There were 9 misclassifications and all of them were classified with less than 15\% confidence by the classifier.

Where $V_j(i) = 1$ if the $j^{th}$ tree voted for class $i$ and $0$
otherwise, and $|T|_f$ denotes the total number of trees in forest $f$.
Note that by construction, $\sum_i P(C_i) = 1$ and $P(C_i)\geq 0$ $\forall$ $i$, allowing us to treat $P(C_i)$ as a probability measure.

%There was one correct classification made with 13.7\% confidence. This suggests that we can use a threshold between 13.7\% and 15\% confidence level for verification, and manually analyze the classifications that did not pass the confidence threshold or exclude them from results.
%
%We picked an aggressive threshold of 15\% and to validate it, we trained a random forest classifier on the same set of 30 programmers 270 code samples. We tested on 150 different files from the programmers in the training set. There were 6 classifications below the 15\% threshold and two of them were misclassified. We took another set of 420 test files from 30 programmers that were not in the training set. All the files from the 30 programmers were attributed to one of the 30 programmers in the training set since this is a closed world classification task, however, the highest confidence level in these classifications was 14.7\%. The 15\% threshold catches all the instances that do not belong to the programmers in the suspect set, gets rid of 2 misclassifications and 4 correct classifications. Consequently, when we see a classification with less than a threshold value, we can reject the classification and attribute the test instance to an unknown suspect.

%
%\begin{figure}[htbp]
%\begin{center}
% \includegraphics[width=0.49\textwidth]{figs/marginCurve.png}
%\caption{ Confidence Thresholds from Classification Margin Curves for Classification Verification}
%\label{fig:closed}
%\end{center}
%\end{figure}

\subsection{\textbf{Case Study: Nulled.IO Hacker Forum}}
On May 6, 2016 the well known `hacker' forum \emph{Nulled.IO} was compromised and its forum dump was leaked along with the private messages of its 585,897 members. The members of these forums share, sell, and buy stolen credentials and cracking software. A high number of the forum members are active developers that write their own code and sell them, or share some of their code for free in public GitHub repositories along with tutorials on how to use them. The private messages of the sellers in the forum include links to their products and even to screenshots of how the products work, for buyers. We were able to find declared authorship along with active links to members' software on sharing sites such as FileDropper\footnote{www.filedropper.com: `Simplest File Hosting Website..'} and MediaFire\footnote{www.mediafire.com: `All your media, anywhere you go'} in the private messages. 

For our case study, we created a dataset from four forum members with a total of thirteen Windows executables. One of the members had only one sample, which we used to test the open world setting described in Section~\ref{sec:openworld}. A challenge encountered in this case study is that the binary programs obtained from Nulled.IO do not contain native code, but bytecode for the Microsoft Common Language Infrastructure (CLI). Therefore, we cannot immediately analyze them using our existing
toolchain. We address this problem by first translating bytecode into corresponding native code using the Microsoft Native Image Generator
(ngen.exe), and subsequently forcing the decompiler to treat the generated output files as regular native code for binaries. On the other hand, \emph{radare2} is not able to disassemble such output or the original executables. Consequently we had access to a subset of the information gain feature set obtained from GCJ. We extracted a total of 605 features consisting of decompiled source code features and \emph{ndisasm} disassembly features. Nevertheless, we are able to de-anonymize these programmers with 100\% accuracy while the one sample from the open world class is classified in all cases with the lowest confidence, such as 0.4, which is below the verification threshold and is recognized by the classifier as a sample that does not belong to the rest of the programmers.

A larger de-anonymization attack can be carried out by collecting code from GitHub users with relevant repositories and identifying all the available executables mentioned in the public portions of hacker forums. GitHub code can be compiled with necessary parameters and used with the approach described in Section~\ref{sec:gitExp}. Incorporating verification thresholds from Section~\ref{sec:openworld} can help handle programmers with only one sample. Consequently a large number of members can be linked, reduced to a cluster or directly de-anonymized.

The countermeasure against real-world programmer de-anonymization attacks requires a combination of various precautions. Developers should not have any public repositories. A set of programs should not be released by the same online identity. Programmers should try to have a different coding style in each piece of software they write and also try to code in different programming languages. Software should utilize different optimizations and obfuscations to avoid deterministic patterns. A programmer who accomplishes randomness across all potential identifying factors would be very difficult to de-anonymize. Nevertheless, even the most privacy savvy developer might be willing to contribute to open source software or build a reputation for her identity based on her set of products, which would be a challenge for maintaining anonymity.

Some of these developers obfuscate their code with the primary goal of hiding the source code and consequently they are experienced in writing or using obfuscators and deobfuscators. 
%These members belong to groups in the forum such as VIP, contributor, reverser, and legendary reverser.
%4 authors from nulled IO with 13 samples. Extracted 600 IG features from ndisasm disassembly and scaa features. 1st one has 1 sample, 2nd one has 7 samples, 3rd one has 3 samples, and the 4th one has 2 samples. 
%\\Train on: 2 samples from class 4, 3 samples from class 2, and 3 samples from class 3. Test on 1 sample from class one and 4 samples from class 2. All class 2 apptributed correctyl. The open world instance is attributed with 0.404, which is lower than the open world threshold. All the correct classifications have higher confidence. 
%\\1,4,2 samples from classes 4,2,3 respectively. Test on one instance from class 1, Open class instance classified with 0.2 threshold, and the rest is correct and above 0.6 confidence.
%\\Small set because hard to find samples for which the authors confessed that they belong to them. The link to most of the other samples is not active anymore (sharing over cloud services after payment). Could be semi-automated in the future. But it is worth noting that decompiling these is cumbersome. And radare does not work on these.
%These people (800K) have github repos but not in c and c++ for the ones that we have forum code samples from. They usually sell these in the forum. The public github code is usually malicious but free. They can probably be linked from their git code's similarity to forum code.
%\\
%\aylin{Give anectodal examples from member writings\\}
An additional challenge encountered in this case study is that the
binary programs obtained from NULLED.io do not contain native code, but
bytecode for the Microsoft Common Language Infrastructure (CLI).
Therefore, we cannot immediately analyze them using our existing
toolchain. We address this problem by first translating bytecode into
corresponding native code using the Microsoft Native Image Generator
(ngen), and subsequently forcing the decompiler to treat the generated
output files as regular native code binaries.

%% file: 4_data.tex
\subsection{\textbf{Dataset}}
\label{sec:dataset}
We evaluate our executable binary authorship attribution method on a controlled dataset based on the annual programming competition \emph{GCJ} \cite{gcj}. It is an annual contest that thousands of programmers take part in each year, including professionals, students, and hobbyists from all over the world. The contestants implement solutions to the same tasks in a limited amount of time in a programming language of their choice. Accordingly, all the correct solutions have the same algorithmic functionality.
There are two main reasons for choosing GCJ competition solutions as an evaluation corpus. First, it enables us to directly compare our results to previous work on executable binary authorship attribution as both \cite{alrabaee2014oba2} and \cite{rosenblum2011wrote} evaluate their approaches on data from GCJ. Second, we eliminate the potential confounding effect of identifying programming task rather than programmer by identifying functionality properties instead of stylistic properties. GCJ is a less noisy and clean dataset known definitely to be single authored. GCJ solutions do not have significant dependencies outside of the standard library and contain few or no third party libraries. 

We focus our analysis on compiled C++ code, the most popular programming language used in the competition. We collect the solutions from the years 2008 to 2014 along with author names and problem identifiers. In GCJ experiments we are assuming that the programmers are not deliberately trying to hide their identity. Accordingly, we show results without excluding symbol information.

\subsection{\textbf{Code Compilation}}

To create our experimental datasets, we first compiled the source code with GNU Compiler Collection's gcc or g++ without any optimization to Executable and Linkable Format (ELF) 32-bit, Intel 80386 Unix binaries. The training set needs to be compiled with the same compiler and settings otherwise we might end up detecting the compiler instead of the author. Passing the training samples through the same encoder preserves mutual information between code style and labels and accordingly we can successfully de-anonymize programmers.

Next, to measure the effect of different compilation options, such as compiler optimization flags, we additionally compiled the source code with level-1, level-2, and level-3 optimizations, namely the O1, O2, and O3 flags. O3 is a superset of O2 optimization flags and similarly O2 is a superset of O1 flags. The compiler attempts to improve the performance and/or code size when the compiler flags are turned on but at the same time optimization has the expense of increasing compilation time and complicating program debugging.

%% file: reconstructedFeatures.tex
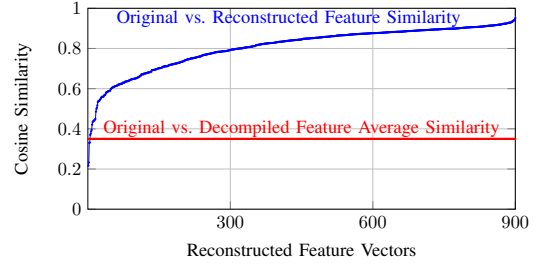
\begin{figure}
\begin{center}
\begin{tikzpicture}[node distance=8pt and 33pt, scale=0.7]
	\begin{axis}[
	height=5.4cm,
	width=9.7cm,
		xlabel=Reconstructed Feature Vectors,
				ylabel=Cosine Similarity,
		 xtick={ 300, 600,900},
		 ymin =0, ymax=1,
		 xmin=0, xmax=900,
		 grid=both,
    grid style={line width=.1pt, draw=gray!10},
    major grid style={line width=.2pt,draw=gray!50},
    ]
			\addplot[color=blue,mark=*,mark size=0.3pt] coordinates {
	%	\addplot[color=blue] coordinates {
(1,0.216576727405941)
(2,0.232554385340787)
(3,0.330535931549103)
(4,0.331260824509672)
(5,0.369643975322497)
(6,0.375892762762694)
(7,0.387624517902651)
(8,0.394421765782602)
(9,0.420878554499445)
(10,0.434187228771628)
(11,0.436475545635043)
(12,0.439057628809777)
(13,0.441815898168268)
(14,0.442553878754297)
(15,0.446525042340284)
(16,0.449641078975954)
(17,0.485643789747317)
(18,0.498880493440964)
(19,0.505415611854027)
(20,0.516926425412207)
(21,0.532890798004531)
(22,0.537698374197557)
(23,0.538945501096215)
(24,0.538977306139326)
(25,0.543787443306089)
(26,0.546509679177614)
(27,0.549955812214659)
(28,0.553368291899548)
(29,0.559124937644815)
(30,0.559291086057741)
(31,0.559607909250483)
(32,0.559902647335853)
(33,0.560816383447162)
(34,0.565231280577114)
(35,0.567750792576144)
(36,0.569277150187936)
(37,0.570240804188626)
(38,0.578317875102201)
(39,0.581726311712125)
(40,0.582182619823015)
(41,0.58440829211762)
(42,0.586423704390902)
(43,0.586548440523272)
(44,0.587323606549334)
(45,0.588727669307633)
(46,0.589152199487554)
(47,0.590269033610648)
(48,0.597847258045194)
(49,0.599111209033695)
(50,0.602029033707109)
(51,0.604700861459208)
(52,0.604941914362753)
(53,0.606713973313724)
(54,0.608222464399153)
(55,0.608582803105819)
(56,0.611241469948105)
(57,0.611939091851649)
(58,0.612467575350428)
(59,0.613193940105219)
(60,0.613732264564045)
(61,0.61490882680877)
(62,0.616348284086566)
(63,0.616689645628831)
(64,0.619998616484425)
(65,0.620276601483372)
(66,0.620976953417321)
(67,0.621086047734507)
(68,0.623487771952516)
(69,0.623648652798534)
(70,0.624308423431297)
(71,0.624588357644431)
(72,0.62461359161991)
(73,0.629225841729597)
(74,0.629537131151785)
(75,0.629660428107189)
(76,0.629953123421903)
(77,0.630568331241602)
(78,0.633861515724617)
(79,0.635078169637135)
(80,0.635234327946458)
(81,0.635437359871912)
(82,0.636782960567915)
(83,0.636967919432409)
(84,0.637423778861961)
(85,0.637527174102839)
(86,0.638379244493273)
(87,0.641546183239895)
(88,0.641564327395766)
(89,0.643444410163278)
(90,0.644134501224167)
(91,0.645822495286164)
(92,0.646277105266292)
(93,0.64680413973407)
(94,0.647313646653101)
(95,0.648477521128095)
(96,0.648562953010499)
(97,0.649031180250837)
(98,0.649497117108292)
(99,0.649933104246014)
(100,0.650356371078531)
(101,0.651139796534563)
(102,0.652440309320017)
(103,0.65314449879371)
(104,0.654204455349813)
(105,0.655844880265198)
(106,0.655970500124433)
(107,0.657934619227049)
(108,0.659206338161907)
(109,0.661117107230736)
(110,0.663117394217278)
(111,0.663333237583589)
(112,0.667720919652172)
(113,0.670838113265962)
(114,0.670997226362822)
(115,0.671262407370573)
(116,0.671322311216345)
(117,0.671626174251653)
(118,0.674840756671605)
(119,0.674917332988645)
(120,0.675135682843317)
(121,0.675634533027371)
(122,0.676337554651356)
(123,0.676857778528432)
(124,0.678346619299761)
(125,0.678800500443923)
(126,0.679288163702137)
(127,0.680621016646625)
(128,0.680701967757377)
(129,0.681252282071912)
(130,0.68154688713964)
(131,0.682861693210699)
(132,0.683194916526555)
(133,0.685103884218785)
(134,0.685586142179163)
(135,0.691969354591524)
(136,0.692823244578917)
(137,0.692865396222044)
(138,0.69323030688245)
(139,0.6932704396059)
(140,0.693810960086014)
(141,0.694281146387796)
(142,0.694534489340089)
(143,0.696319356238725)
(144,0.696928867679272)
(145,0.697129816917072)
(146,0.697713666189231)
(147,0.699128307664364)
(148,0.700720142300344)
(149,0.701290287653874)
(150,0.701651431971875)
(151,0.701784658446239)
(152,0.702290027389768)
(153,0.702533463388501)
(154,0.703312473289982)
(155,0.703795035151396)
(156,0.705121039918056)
(157,0.706813441220704)
(158,0.708995105290137)
(159,0.709324786072429)
(160,0.70932677519445)
(161,0.710893955050005)
(162,0.710907505592573)
(163,0.710979784324856)
(164,0.711337912862333)
(165,0.711498793889974)
(166,0.711548712690905)
(167,0.711613798261338)
(168,0.713036571230208)
(169,0.714256869578546)
(170,0.715365233793803)
(171,0.716315867916244)
(172,0.718093087059964)
(173,0.718126504755401)
(174,0.718971344962238)
(175,0.719682078510573)
(176,0.719751045526723)
(177,0.720204607439211)
(178,0.720972909820815)
(179,0.721075392416441)
(180,0.721846994700709)
(181,0.722433329325331)
(182,0.72352850582995)
(183,0.723758654954404)
(184,0.724813924584139)
(185,0.725064904305451)
(186,0.725146602348971)
(187,0.726066126769407)
(188,0.726133261073234)
(189,0.7271569465313)
(190,0.728595799002873)
(191,0.728694514409233)
(192,0.729249306082602)
(193,0.729751200105919)
(194,0.730254724955656)
(195,0.731919598517434)
(196,0.73275480333535)
(197,0.73374249667686)
(198,0.733785814266215)
(199,0.734267913052025)
(200,0.735390915181347)
(201,0.736457483317974)
(202,0.737003443296723)
(203,0.739421451946312)
(204,0.740669285319082)
(205,0.741046579761264)
(206,0.741704315929274)
(207,0.743315997533082)
(208,0.743892218861416)
(209,0.745127754253704)
(210,0.745323230841713)
(211,0.745680629895705)
(212,0.746253548875239)
(213,0.746840793570876)
(214,0.747525376264472)
(215,0.748550661412603)
(216,0.749105237245523)
(217,0.750277876325305)
(218,0.75053347553996)
(219,0.75079824702308)
(220,0.751054537519409)
(221,0.751137917482094)
(222,0.752007530352369)
(223,0.752299458180376)
(224,0.753641757084745)
(225,0.753666669849031)
(226,0.753672292243678)
(227,0.754425308210137)
(228,0.755528501970366)
(229,0.756115915525723)
(230,0.756846350329255)
(231,0.757765585188247)
(232,0.757987227196869)
(233,0.758066960290845)
(234,0.759876205354878)
(235,0.760035874863354)
(236,0.760167936571269)
(237,0.760177225679079)
(238,0.76055557759385)
(239,0.76274138433292)
(240,0.764310200510026)
(241,0.765247005627585)
(242,0.765336569662864)
(243,0.76664715863978)
(244,0.767670883688084)
(245,0.768427326207237)
(246,0.768652003643117)
(247,0.768987737324112)
(248,0.769484349666668)
(249,0.769755298910971)
(250,0.770324336866028)
(251,0.770600402382935)
(252,0.771784639506559)
(253,0.77202502322563)
(254,0.77242860330889)
(255,0.773170133405138)
(256,0.77328827654281)
(257,0.773924738925666)
(258,0.774702685077293)
(259,0.775483566876542)
(260,0.775517859307984)
(261,0.776038847125971)
(262,0.776911200850784)
(263,0.777095579381832)
(264,0.777611825996406)
(265,0.778646251734112)
(266,0.779396581363807)
(267,0.779554977737969)
(268,0.780955438576135)
(269,0.780970219859646)
(270,0.781448515443663)
(271,0.781573906276622)
(272,0.781808706161292)
(273,0.78273145880129)
(274,0.782961578409783)
(275,0.783458166771246)
(276,0.78376286751842)
(277,0.783800248966809)
(278,0.78381933869314)
(279,0.783837787038889)
(280,0.784147626075914)
(281,0.784844811668585)
(282,0.784928778562181)
(283,0.785248590721517)
(284,0.785777431518756)
(285,0.78620295881264)
(286,0.786277998377639)
(287,0.786382701046627)
(288,0.786471117536779)
(289,0.786557426825192)
(290,0.786611534479242)
(291,0.787156141501457)
(292,0.787689713955107)
(293,0.788471274103428)
(294,0.788566284054158)
(295,0.789167409243684)
(296,0.789384013500342)
(297,0.79035319335797)
(298,0.791212010219644)
(299,0.791425286996157)
(300,0.792748589162314)
(301,0.793456514075123)
(302,0.793594987330192)
(303,0.794768260991468)
(304,0.79481509265881)
(305,0.795186728017361)
(306,0.795222729035307)
(307,0.795336408935463)
(308,0.795449233734052)
(309,0.796310483227486)
(310,0.796891265246024)
(311,0.797411594277158)
(312,0.797628382792252)
(313,0.798100227541081)
(314,0.798404132599036)
(315,0.798639538777863)
(316,0.798848116912194)
(317,0.799293447312777)
(318,0.799534754307731)
(319,0.799883090869163)
(320,0.800199922972563)
(321,0.800290368568015)
(322,0.800294800419712)
(323,0.800324211892025)
(324,0.800874637381414)
(325,0.801112373337571)
(326,0.801785596544273)
(327,0.801923360444822)
(328,0.803360199890128)
(329,0.803615669669794)
(330,0.803680515778577)
(331,0.804553985068184)
(332,0.804680561351362)
(333,0.804693812338465)
(334,0.804914456324867)
(335,0.806552508328042)
(336,0.806636658802476)
(337,0.806685404599016)
(338,0.806776034748798)
(339,0.806819869387181)
(340,0.807415215874661)
(341,0.80780056804864)
(342,0.808302287872197)
(343,0.808458968719142)
(344,0.808973031563901)
(345,0.809558042940983)
(346,0.809590587257916)
(347,0.809728213624392)
(348,0.811538776961692)
(349,0.811702233435758)
(350,0.811817801608399)
(351,0.811881189686391)
(352,0.811968039721385)
(353,0.814775797897083)
(354,0.815747614845567)
(355,0.816535442873699)
(356,0.816555851219796)
(357,0.816728311574128)
(358,0.817844679488729)
(359,0.818339517540901)
(360,0.818964297910873)
(361,0.819508900054168)
(362,0.819625273592524)
(363,0.819641157409146)
(364,0.820168640761151)
(365,0.820986413298669)
(366,0.821590605387042)
(367,0.821895843280972)
(368,0.822017790698904)
(369,0.822342121223323)
(370,0.822731738808886)
(371,0.823095770892905)
(372,0.823763630479878)
(373,0.823890029867542)
(374,0.823906549778102)
(375,0.824389765114251)
(376,0.824661319775999)
(377,0.825171058730446)
(378,0.825239323025949)
(379,0.825687296098631)
(380,0.825735600798973)
(381,0.825984524798555)
(382,0.826569732707994)
(383,0.826723442668749)
(384,0.826925706085222)
(385,0.827027188171418)
(386,0.827107112986536)
(387,0.827161507933454)
(388,0.827333767163565)
(389,0.827336960603967)
(390,0.827564550225755)
(391,0.82764379629042)
(392,0.828069950746775)
(393,0.828961044058552)
(394,0.829196939777439)
(395,0.829855789923145)
(396,0.830403500192461)
(397,0.83058057256391)
(398,0.83080385385951)
(399,0.831538243147316)
(400,0.83203924395466)
(401,0.832235937604776)
(402,0.832761027180529)
(403,0.832905732988374)
(404,0.833336246170997)
(405,0.833344644884438)
(406,0.833577322009148)
(407,0.833609649446929)
(408,0.833697237659733)
(409,0.83431580487218)
(410,0.834744640247282)
(411,0.835186984432766)
(412,0.835380012748804)
(413,0.835432377959474)
(414,0.835626212195063)
(415,0.835768546333542)
(416,0.836636966659144)
(417,0.837161528854051)
(418,0.837211675675631)
(419,0.837841406351349)
(420,0.837848248542639)
(421,0.837872706428567)
(422,0.837969500534614)
(423,0.838493868664173)
(424,0.839167273036962)
(425,0.839192238106411)
(426,0.839529504696826)
(427,0.839618333810067)
(428,0.839841555675864)
(429,0.839895694884309)
(430,0.840529925284774)
(431,0.840797790063312)
(432,0.8412099160585)
(433,0.841249864848287)
(434,0.841280505793519)
(435,0.841302778154945)
(436,0.841998970790851)
(437,0.842099459984902)
(438,0.84245337222238)
(439,0.842517544805638)
(440,0.842559084494322)
(441,0.842681741973894)
(442,0.842763138371324)
(443,0.843430926458128)
(444,0.843511058918205)
(445,0.843930469315359)
(446,0.843984556030975)
(447,0.843987395847195)
(448,0.844028471203008)
(449,0.844647434496878)
(450,0.84567035812098)
(451,0.84624288653876)
(452,0.846311517003683)
(453,0.847255475048562)
(454,0.84837145726529)
(455,0.848423276782414)
(456,0.848617671260013)
(457,0.848634717668682)
(458,0.848686447448035)
(459,0.849112760838039)
(460,0.849124067678551)
(461,0.849199875121307)
(462,0.849302488980488)
(463,0.849330249528089)
(464,0.84936312967081)
(465,0.849605350550632)
(466,0.849608864820425)
(467,0.850330446328204)
(468,0.850672280140528)
(469,0.850929140161592)
(470,0.85111400263665)
(471,0.851126205376495)
(472,0.851380653866985)
(473,0.851485195716323)
(474,0.85151272666599)
(475,0.851535166642354)
(476,0.851847348672967)
(477,0.852438601051857)
(478,0.852562960293451)
(479,0.852576630820626)
(480,0.853250042491813)
(481,0.853332941658267)
(482,0.853479948819665)
(483,0.853538508409872)
(484,0.853626846960148)
(485,0.853633023537636)
(486,0.854194411428874)
(487,0.854690276979641)
(488,0.85470098510182)
(489,0.854752300522228)
(490,0.855062137964127)
(491,0.855489447203162)
(492,0.855595783751055)
(493,0.855692603200013)
(494,0.856632032213951)
(495,0.856826073706368)
(496,0.856989026796019)
(497,0.857071522279721)
(498,0.857593329457036)
(499,0.857612449291966)
(500,0.857644326888534)
(501,0.857923416596862)
(502,0.858098148703128)
(503,0.858115668718053)
(504,0.858248813433952)
(505,0.858435157137151)
(506,0.858661891266507)
(507,0.858876468112511)
(508,0.859300995456923)
(509,0.85944419258812)
(510,0.860024963542968)
(511,0.86017475432495)
(512,0.860471548880796)
(513,0.860949384171607)
(514,0.861066095378939)
(515,0.861101485770701)
(516,0.861256014331451)
(517,0.861268668649111)
(518,0.861468602859369)
(519,0.861653165165262)
(520,0.861728083923183)
(521,0.862086777444853)
(522,0.862148453685826)
(523,0.86288739279961)
(524,0.864022013736618)
(525,0.864284643539346)
(526,0.864341100836227)
(527,0.864405289046653)
(528,0.864419286860567)
(529,0.864877932526816)
(530,0.865137925360166)
(531,0.865257632445812)
(532,0.865330768365861)
(533,0.865482896041313)
(534,0.865697844105884)
(535,0.865818316970732)
(536,0.865827401896392)
(537,0.865962176698752)
(538,0.866060203585015)
(539,0.866078117409539)
(540,0.866235137029843)
(541,0.866526147988895)
(542,0.866884330428852)
(543,0.866891006248674)
(544,0.866991047801831)
(545,0.867082810566798)
(546,0.867343794125076)
(547,0.867645059295126)
(548,0.867890297888029)
(549,0.868171735953351)
(550,0.868249499288603)
(551,0.868650794639923)
(552,0.869178818572801)
(553,0.869368428149994)
(554,0.869377867455617)
(555,0.869422805682379)
(556,0.869754132339286)
(557,0.869933636478606)
(558,0.869964451540001)
(559,0.870118797529102)
(560,0.870728814287212)
(561,0.870807961235485)
(562,0.870970459200477)
(563,0.870971272122797)
(564,0.871007740543245)
(565,0.871256150700984)
(566,0.871288490850062)
(567,0.871321795354848)
(568,0.871792715928015)
(569,0.872052539252015)
(570,0.872053086547297)
(571,0.872112346275471)
(572,0.872525305519988)
(573,0.872696358906263)
(574,0.872801413987643)
(575,0.872910421936843)
(576,0.87300266919597)
(577,0.873106019965262)
(578,0.87325681379752)
(579,0.873322122329023)
(580,0.873392874521848)
(581,0.873639472251139)
(582,0.873699581954284)
(583,0.873711413315135)
(584,0.873839037812429)
(585,0.873952811688491)
(586,0.874023763275206)
(587,0.874269746752408)
(588,0.874358788293798)
(589,0.874409197229201)
(590,0.874851434884866)
(591,0.874896267754231)
(592,0.875126486557998)
(593,0.875277195346713)
(594,0.875455144338439)
(595,0.875527157045696)
(596,0.875597159768908)
(597,0.876025324762254)
(598,0.876249073379104)
(599,0.876281012757713)
(600,0.876281898027511)
(601,0.876349566674106)
(602,0.876356278725963)
(603,0.87641366875671)
(604,0.876569307973727)
(605,0.876604312597127)
(606,0.876718787362879)
(607,0.876785318389449)
(608,0.876985698486554)
(609,0.877013841275235)
(610,0.877087846776684)
(611,0.877198380364682)
(612,0.877338562121743)
(613,0.877472723545882)
(614,0.877530395415897)
(615,0.877727981920223)
(616,0.877741527728208)
(617,0.877882529678126)
(618,0.877936673104892)
(619,0.878073731406876)
(620,0.878320240451687)
(621,0.878539425544285)
(622,0.878775220272078)
(623,0.878845805261019)
(624,0.878902516777934)
(625,0.879084396156407)
(626,0.879165655602885)
(627,0.879552009573399)
(628,0.879919673141236)
(629,0.880202867759008)
(630,0.880348836455946)
(631,0.880395295236437)
(632,0.880512161810672)
(633,0.880531802136147)
(634,0.880740699458501)
(635,0.881190525297727)
(636,0.881463634619494)
(637,0.881518317535769)
(638,0.881627633944267)
(639,0.881748961619587)
(640,0.881872333910644)
(641,0.881972601729724)
(642,0.88203660599414)
(643,0.882039316443369)
(644,0.882051612776666)
(645,0.882095235562668)
(646,0.882197742126936)
(647,0.88238537245304)
(648,0.882556413406262)
(649,0.882753701459584)
(650,0.88303927762434)
(651,0.883106815474316)
(652,0.883108022920093)
(653,0.883128883926031)
(654,0.883226305066991)
(655,0.883254956863019)
(656,0.883284634827239)
(657,0.883388663815146)
(658,0.88352138979695)
(659,0.88382389734415)
(660,0.883902837112003)
(661,0.883952899015261)
(662,0.883974391293861)
(663,0.884065723467588)
(664,0.884446564806637)
(665,0.884610176760425)
(666,0.884974861112325)
(667,0.885157489335004)
(668,0.885168120157605)
(669,0.885348078326755)
(670,0.885600682418929)
(671,0.885867888351164)
(672,0.886141842309755)
(673,0.886449817780729)
(674,0.886656049493368)
(675,0.886708193154824)
(676,0.886846761374213)
(677,0.886853053956294)
(678,0.886975221628124)
(679,0.887044467089477)
(680,0.887065000740667)
(681,0.887119850671502)
(682,0.887203942972868)
(683,0.887506311991057)
(684,0.88752024146806)
(685,0.887642775455825)
(686,0.887864161729802)
(687,0.888174969918529)
(688,0.888259761993404)
(689,0.888268674502439)
(690,0.888276688265942)
(691,0.888674042140333)
(692,0.888961614958044)
(693,0.889224784564415)
(694,0.889262515169167)
(695,0.889334677143168)
(696,0.889457204408515)
(697,0.890038990640555)
(698,0.890053727494861)
(699,0.890099298390831)
(700,0.890233542195731)
(701,0.89058846591198)
(702,0.89064833924567)
(703,0.890769970469808)
(704,0.890919317568668)
(705,0.890978181406295)
(706,0.891053254061888)
(707,0.891119870377042)
(708,0.891214934723479)
(709,0.891217067574951)
(710,0.891380044620768)
(711,0.891432525481361)
(712,0.891713032202817)
(713,0.891889486679285)
(714,0.892159753002254)
(715,0.892347385266114)
(716,0.892367278145382)
(717,0.892469221055504)
(718,0.892477334424133)
(719,0.892584643747175)
(720,0.892792051988661)
(721,0.893341041319447)
(722,0.893357798368736)
(723,0.893598862249977)
(724,0.893625699106877)
(725,0.893714341505277)
(726,0.893848683187076)
(727,0.893967829464603)
(728,0.89410023355303)
(729,0.894241473809378)
(730,0.894286623655045)
(731,0.894468546420018)
(732,0.894497310469266)
(733,0.894573818108532)
(734,0.894666894190982)
(735,0.894718740477092)
(736,0.894727390190551)
(737,0.894935046698451)
(738,0.895006452027077)
(739,0.895395691009343)
(740,0.895644527040273)
(741,0.895839618077795)
(742,0.895842675630965)
(743,0.895981766550123)
(744,0.896022204146745)
(745,0.896260233209802)
(746,0.896395563491326)
(747,0.896806566961384)
(748,0.896850400636692)
(749,0.896890752379998)
(750,0.897001481271492)
(751,0.897039943620972)
(752,0.897100165675485)
(753,0.897141798629964)
(754,0.897624095230586)
(755,0.897714787286432)
(756,0.897731309939454)
(757,0.897887937967813)
(758,0.89793117224472)
(759,0.89793622100268)
(760,0.898080557874111)
(761,0.898104676035207)
(762,0.898413783797937)
(763,0.898420598531837)
(764,0.898448910494362)
(765,0.898570117628415)
(766,0.898608205347785)
(767,0.898770379422969)
(768,0.898888560031871)
(769,0.898903636845984)
(770,0.898929231722084)
(771,0.898969306047852)
(772,0.899045465109775)
(773,0.899164986142546)
(774,0.899215774244159)
(775,0.899254903125736)
(776,0.900011664623622)
(777,0.900230762946779)
(778,0.900321084107784)
(779,0.90034943814961)
(780,0.900462606051752)
(781,0.900798834602556)
(782,0.900987694327221)
(783,0.901094512475281)
(784,0.901447599383032)
(785,0.901806333378968)
(786,0.901877368636128)
(787,0.902043186803422)
(788,0.902335110894851)
(789,0.902471999979323)
(790,0.902721539029328)
(791,0.902952847268215)
(792,0.903283199078397)
(793,0.903479281641564)
(794,0.903580471359066)
(795,0.903640193436252)
(796,0.904024894918634)
(797,0.90451801559473)
(798,0.904612751916543)
(799,0.904709376243342)
(800,0.904749650588704)
(801,0.904931122853223)
(802,0.905044063685361)
(803,0.905048718072654)
(804,0.905301274941267)
(805,0.905575769341727)
(806,0.90558198038101)
(807,0.9058577635881)
(808,0.905941298193776)
(809,0.906048528512503)
(810,0.906105655869496)
(811,0.90636603783914)
(812,0.906374799301059)
(813,0.906477018110425)
(814,0.906480151222752)
(815,0.906720006270075)
(816,0.906864560314388)
(817,0.906870604833639)
(818,0.906886194372294)
(819,0.907193286205286)
(820,0.907228325630457)
(821,0.907372432895963)
(822,0.907745475268559)
(823,0.907761844009473)
(824,0.90801263541067)
(825,0.908052703545115)
(826,0.908704570790145)
(827,0.908743606520467)
(828,0.908921642579767)
(829,0.908945222443142)
(830,0.909057712166037)
(831,0.909119878773064)
(832,0.90954758657336)
(833,0.909695861170437)
(834,0.910228465783539)
(835,0.910502927684903)
(836,0.910674102938695)
(837,0.911141232405638)
(838,0.911325521122118)
(839,0.911564367757147)
(840,0.911681204694122)
(841,0.911831383746843)
(842,0.912667538731821)
(843,0.912934072829666)
(844,0.913125643125849)
(845,0.913349963522605)
(846,0.913716169587782)
(847,0.913813777012116)
(848,0.913865449160105)
(849,0.913990015877561)
(850,0.914019596110174)
(851,0.914304458085401)
(852,0.914370909649542)
(853,0.914468199927398)
(854,0.914659105763068)
(855,0.915272872864976)
(856,0.915721655175433)
(857,0.915816186423146)
(858,0.916121166507485)
(859,0.916303370325443)
(860,0.916444982171089)
(861,0.916533914450402)
(862,0.916971511104505)
(863,0.917422093296977)
(864,0.918162562043647)
(865,0.918511307578947)
(866,0.918640342688952)
(867,0.918650495918166)
(868,0.918692762229371)
(869,0.919153963306504)
(870,0.919974335488606)
(871,0.920246179942452)
(872,0.920926201770028)
(873,0.92203660921739)
(874,0.922283235050738)
(875,0.922577065214686)
(876,0.923008386370047)
(877,0.923439284469497)
(878,0.923444339709764)
(879,0.923782798407037)
(880,0.923986187305168)
(881,0.924075680324065)
(882,0.924599119514968)
(883,0.926176494236046)
(884,0.926196336809201)
(885,0.927191012066478)
(886,0.92780638742534)
(887,0.928155577625922)
(888,0.928736541286055)
(889,0.928827108064879)
(890,0.929186217541368)
(891,0.929857840694389)
(892,0.932887349348965)
(893,0.932958173569089)
(894,0.933205073521217)
(895,0.936648278959115)
(896,0.938839811714252)
(897,0.939166985063259)
(898,0.94177397367544)
(899,0.945578741268875)
(900,0.951314806358503)
			};
	
	\addplot[color=red, mark size=0.3pt,
                mark=*] coordinates {
	(1,0.35)
(2,0.35)
(3,0.35)
(4,0.35)
(5,0.35)
(6,0.35)
(7,0.35)
(8,0.35)
(9,0.35)
(10,0.35)
(11,0.35)
(12,0.35)
(13,0.35)
(14,0.35)
(15,0.35)
(16,0.35)
(17,0.35)
(18,0.35)
(19,0.35)
(20,0.35)
(21,0.35)
(22,0.35)
(23,0.35)
(24,0.35)
(25,0.35)
(26,0.35)
(27,0.35)
(28,0.35)
(29,0.35)
(30,0.35)
(31,0.35)
(32,0.35)
(33,0.35)
(34,0.35)
(35,0.35)
(36,0.35)
(37,0.35)
(38,0.35)
(39,0.35)
(40,0.35)
(41,0.35)
(42,0.35)
(43,0.35)
(44,0.35)
(45,0.35)
(46,0.35)
(47,0.35)
(48,0.35)
(49,0.35)
(50,0.35)
(51,0.35)
(52,0.35)
(53,0.35)
(54,0.35)
(55,0.35)
(56,0.35)
(57,0.35)
(58,0.35)
(59,0.35)
(60,0.35)
(61,0.35)
(62,0.35)
(63,0.35)
(64,0.35)
(65,0.35)
(66,0.35)
(67,0.35)
(68,0.35)
(69,0.35)
(70,0.35)
(71,0.35)
(72,0.35)
(73,0.35)
(74,0.35)
(75,0.35)
(76,0.35)
(77,0.35)
(78,0.35)
(79,0.35)
(80,0.35)
(81,0.35)
(82,0.35)
(83,0.35)
(84,0.35)
(85,0.35)
(86,0.35)
(87,0.35)
(88,0.35)
(89,0.35)
(90,0.35)
(91,0.35)
(92,0.35)
(93,0.35)
(94,0.35)
(95,0.35)
(96,0.35)
(97,0.35)
(98,0.35)
(99,0.35)
(100,0.35)
(101,0.35)
(102,0.35)
(103,0.35)
(104,0.35)
(105,0.35)
(106,0.35)
(107,0.35)
(108,0.35)
(109,0.35)
(110,0.35)
(111,0.35)
(112,0.35)
(113,0.35)
(114,0.35)
(115,0.35)
(116,0.35)
(117,0.35)
(118,0.35)
(119,0.35)
(120,0.35)
(121,0.35)
(122,0.35)
(123,0.35)
(124,0.35)
(125,0.35)
(126,0.35)
(127,0.35)
(128,0.35)
(129,0.35)
(130,0.35)
(131,0.35)
(132,0.35)
(133,0.35)
(134,0.35)
(135,0.35)
(136,0.35)
(137,0.35)
(138,0.35)
(139,0.35)
(140,0.35)
(141,0.35)
(142,0.35)
(143,0.35)
(144,0.35)
(145,0.35)
(146,0.35)
(147,0.35)
(148,0.35)
(149,0.35)
(150,0.35)
(151,0.35)
(152,0.35)
(153,0.35)
(154,0.35)
(155,0.35)
(156,0.35)
(157,0.35)
(158,0.35)
(159,0.35)
(160,0.35)
(161,0.35)
(162,0.35)
(163,0.35)
(164,0.35)
(165,0.35)
(166,0.35)
(167,0.35)
(168,0.35)
(169,0.35)
(170,0.35)
(171,0.35)
(172,0.35)
(173,0.35)
(174,0.35)
(175,0.35)
(176,0.35)
(177,0.35)
(178,0.35)
(179,0.35)
(180,0.35)
(181,0.35)
(182,0.35)
(183,0.35)
(184,0.35)
(185,0.35)
(186,0.35)
(187,0.35)
(188,0.35)
(189,0.35)
(190,0.35)
(191,0.35)
(192,0.35)
(193,0.35)
(194,0.35)
(195,0.35)
(196,0.35)
(197,0.35)
(198,0.35)
(199,0.35)
(200,0.35)
(201,0.35)
(202,0.35)
(203,0.35)
(204,0.35)
(205,0.35)
(206,0.35)
(207,0.35)
(208,0.35)
(209,0.35)
(210,0.35)
(211,0.35)
(212,0.35)
(213,0.35)
(214,0.35)
(215,0.35)
(216,0.35)
(217,0.35)
(218,0.35)
(219,0.35)
(220,0.35)
(221,0.35)
(222,0.35)
(223,0.35)
(224,0.35)
(225,0.35)
(226,0.35)
(227,0.35)
(228,0.35)
(229,0.35)
(230,0.35)
(231,0.35)
(232,0.35)
(233,0.35)
(234,0.35)
(235,0.35)
(236,0.35)
(237,0.35)
(238,0.35)
(239,0.35)
(240,0.35)
(241,0.35)
(242,0.35)
(243,0.35)
(244,0.35)
(245,0.35)
(246,0.35)
(247,0.35)
(248,0.35)
(249,0.35)
(250,0.35)
(251,0.35)
(252,0.35)
(253,0.35)
(254,0.35)
(255,0.35)
(256,0.35)
(257,0.35)
(258,0.35)
(259,0.35)
(260,0.35)
(261,0.35)
(262,0.35)
(263,0.35)
(264,0.35)
(265,0.35)
(266,0.35)
(267,0.35)
(268,0.35)
(269,0.35)
(270,0.35)
(271,0.35)
(272,0.35)
(273,0.35)
(274,0.35)
(275,0.35)
(276,0.35)
(277,0.35)
(278,0.35)
(279,0.35)
(280,0.35)
(281,0.35)
(282,0.35)
(283,0.35)
(284,0.35)
(285,0.35)
(286,0.35)
(287,0.35)
(288,0.35)
(289,0.35)
(290,0.35)
(291,0.35)
(292,0.35)
(293,0.35)
(294,0.35)
(295,0.35)
(296,0.35)
(297,0.35)
(298,0.35)
(299,0.35)
(300,0.35)
(301,0.35)
(302,0.35)
(303,0.35)
(304,0.35)
(305,0.35)
(306,0.35)
(307,0.35)
(308,0.35)
(309,0.35)
(310,0.35)
(311,0.35)
(312,0.35)
(313,0.35)
(314,0.35)
(315,0.35)
(316,0.35)
(317,0.35)
(318,0.35)
(319,0.35)
(320,0.35)
(321,0.35)
(322,0.35)
(323,0.35)
(324,0.35)
(325,0.35)
(326,0.35)
(327,0.35)
(328,0.35)
(329,0.35)
(330,0.35)
(331,0.35)
(332,0.35)
(333,0.35)
(334,0.35)
(335,0.35)
(336,0.35)
(337,0.35)
(338,0.35)
(339,0.35)
(340,0.35)
(341,0.35)
(342,0.35)
(343,0.35)
(344,0.35)
(345,0.35)
(346,0.35)
(347,0.35)
(348,0.35)
(349,0.35)
(350,0.35)
(351,0.35)
(352,0.35)
(353,0.35)
(354,0.35)
(355,0.35)
(356,0.35)
(357,0.35)
(358,0.35)
(359,0.35)
(360,0.35)
(361,0.35)
(362,0.35)
(363,0.35)
(364,0.35)
(365,0.35)
(366,0.35)
(367,0.35)
(368,0.35)
(369,0.35)
(370,0.35)
(371,0.35)
(372,0.35)
(373,0.35)
(374,0.35)
(375,0.35)
(376,0.35)
(377,0.35)
(378,0.35)
(379,0.35)
(380,0.35)
(381,0.35)
(382,0.35)
(383,0.35)
(384,0.35)
(385,0.35)
(386,0.35)
(387,0.35)
(388,0.35)
(389,0.35)
(390,0.35)
(391,0.35)
(392,0.35)
(393,0.35)
(394,0.35)
(395,0.35)
(396,0.35)
(397,0.35)
(398,0.35)
(399,0.35)
(400,0.35)
(401,0.35)
(402,0.35)
(403,0.35)
(404,0.35)
(405,0.35)
(406,0.35)
(407,0.35)
(408,0.35)
(409,0.35)
(410,0.35)
(411,0.35)
(412,0.35)
(413,0.35)
(414,0.35)
(415,0.35)
(416,0.35)
(417,0.35)
(418,0.35)
(419,0.35)
(420,0.35)
(421,0.35)
(422,0.35)
(423,0.35)
(424,0.35)
(425,0.35)
(426,0.35)
(427,0.35)
(428,0.35)
(429,0.35)
(430,0.35)
(431,0.35)
(432,0.35)
(433,0.35)
(434,0.35)
(435,0.35)
(436,0.35)
(437,0.35)
(438,0.35)
(439,0.35)
(440,0.35)
(441,0.35)
(442,0.35)
(443,0.35)
(444,0.35)
(445,0.35)
(446,0.35)
(447,0.35)
(448,0.35)
(449,0.35)
(450,0.35)
(451,0.35)
(452,0.35)
(453,0.35)
(454,0.35)
(455,0.35)
(456,0.35)
(457,0.35)
(458,0.35)
(459,0.35)
(460,0.35)
(461,0.35)
(462,0.35)
(463,0.35)
(464,0.35)
(465,0.35)
(466,0.35)
(467,0.35)
(468,0.35)
(469,0.35)
(470,0.35)
(471,0.35)
(472,0.35)
(473,0.35)
(474,0.35)
(475,0.35)
(476,0.35)
(477,0.35)
(478,0.35)
(479,0.35)
(480,0.35)
(481,0.35)
(482,0.35)
(483,0.35)
(484,0.35)
(485,0.35)
(486,0.35)
(487,0.35)
(488,0.35)
(489,0.35)
(490,0.35)
(491,0.35)
(492,0.35)
(493,0.35)
(494,0.35)
(495,0.35)
(496,0.35)
(497,0.35)
(498,0.35)
(499,0.35)
(500,0.35)
(501,0.35)
(502,0.35)
(503,0.35)
(504,0.35)
(505,0.35)
(506,0.35)
(507,0.35)
(508,0.35)
(509,0.35)
(510,0.35)
(511,0.35)
(512,0.35)
(513,0.35)
(514,0.35)
(515,0.35)
(516,0.35)
(517,0.35)
(518,0.35)
(519,0.35)
(520,0.35)
(521,0.35)
(522,0.35)
(523,0.35)
(524,0.35)
(525,0.35)
(526,0.35)
(527,0.35)
(528,0.35)
(529,0.35)
(530,0.35)
(531,0.35)
(532,0.35)
(533,0.35)
(534,0.35)
(535,0.35)
(536,0.35)
(537,0.35)
(538,0.35)
(539,0.35)
(540,0.35)
(541,0.35)
(542,0.35)
(543,0.35)
(544,0.35)
(545,0.35)
(546,0.35)
(547,0.35)
(548,0.35)
(549,0.35)
(550,0.35)
(551,0.35)
(552,0.35)
(553,0.35)
(554,0.35)
(555,0.35)
(556,0.35)
(557,0.35)
(558,0.35)
(559,0.35)
(560,0.35)
(561,0.35)
(562,0.35)
(563,0.35)
(564,0.35)
(565,0.35)
(566,0.35)
(567,0.35)
(568,0.35)
(569,0.35)
(570,0.35)
(571,0.35)
(572,0.35)
(573,0.35)
(574,0.35)
(575,0.35)
(576,0.35)
(577,0.35)
(578,0.35)
(579,0.35)
(580,0.35)
(581,0.35)
(582,0.35)
(583,0.35)
(584,0.35)
(585,0.35)
(586,0.35)
(587,0.35)
(588,0.35)
(589,0.35)
(590,0.35)
(591,0.35)
(592,0.35)
(593,0.35)
(594,0.35)
(595,0.35)
(596,0.35)
(597,0.35)
(598,0.35)
(599,0.35)
(600,0.35)
(601,0.35)
(602,0.35)
(603,0.35)
(604,0.35)
(605,0.35)
(606,0.35)
(607,0.35)
(608,0.35)
(609,0.35)
(610,0.35)
(611,0.35)
(612,0.35)
(613,0.35)
(614,0.35)
(615,0.35)
(616,0.35)
(617,0.35)
(618,0.35)
(619,0.35)
(620,0.35)
(621,0.35)
(622,0.35)
(623,0.35)
(624,0.35)
(625,0.35)
(626,0.35)
(627,0.35)
(628,0.35)
(629,0.35)
(630,0.35)
(631,0.35)
(632,0.35)
(633,0.35)
(634,0.35)
(635,0.35)
(636,0.35)
(637,0.35)
(638,0.35)
(639,0.35)
(640,0.35)
(641,0.35)
(642,0.35)
(643,0.35)
(644,0.35)
(645,0.35)
(646,0.35)
(647,0.35)
(648,0.35)
(649,0.35)
(650,0.35)
(651,0.35)
(652,0.35)
(653,0.35)
(654,0.35)
(655,0.35)
(656,0.35)
(657,0.35)
(658,0.35)
(659,0.35)
(660,0.35)
(661,0.35)
(662,0.35)
(663,0.35)
(664,0.35)
(665,0.35)
(666,0.35)
(667,0.35)
(668,0.35)
(669,0.35)
(670,0.35)
(671,0.35)
(672,0.35)
(673,0.35)
(674,0.35)
(675,0.35)
(676,0.35)
(677,0.35)
(678,0.35)
(679,0.35)
(680,0.35)
(681,0.35)
(682,0.35)
(683,0.35)
(684,0.35)
(685,0.35)
(686,0.35)
(687,0.35)
(688,0.35)
(689,0.35)
(690,0.35)
(691,0.35)
(692,0.35)
(693,0.35)
(694,0.35)
(695,0.35)
(696,0.35)
(697,0.35)
(698,0.35)
(699,0.35)
(700,0.35)
(701,0.35)
(702,0.35)
(703,0.35)
(704,0.35)
(705,0.35)
(706,0.35)
(707,0.35)
(708,0.35)
(709,0.35)
(710,0.35)
(711,0.35)
(712,0.35)
(713,0.35)
(714,0.35)
(715,0.35)
(716,0.35)
(717,0.35)
(718,0.35)
(719,0.35)
(720,0.35)
(721,0.35)
(722,0.35)
(723,0.35)
(724,0.35)
(725,0.35)
(726,0.35)
(727,0.35)
(728,0.35)
(729,0.35)
(730,0.35)
(731,0.35)
(732,0.35)
(733,0.35)
(734,0.35)
(735,0.35)
(736,0.35)
(737,0.35)
(738,0.35)
(739,0.35)
(740,0.35)
(741,0.35)
(742,0.35)
(743,0.35)
(744,0.35)
(745,0.35)
(746,0.35)
(747,0.35)
(748,0.35)
(749,0.35)
(750,0.35)
(751,0.35)
(752,0.35)
(753,0.35)
(754,0.35)
(755,0.35)
(756,0.35)
(757,0.35)
(758,0.35)
(759,0.35)
(760,0.35)
(761,0.35)
(762,0.35)
(763,0.35)
(764,0.35)
(765,0.35)
(766,0.35)
(767,0.35)
(768,0.35)
(769,0.35)
(770,0.35)
(771,0.35)
(772,0.35)
(773,0.35)
(774,0.35)
(775,0.35)
(776,0.35)
(777,0.35)
(778,0.35)
(779,0.35)
(780,0.35)
(781,0.35)
(782,0.35)
(783,0.35)
(784,0.35)
(785,0.35)
(786,0.35)
(787,0.35)
(788,0.35)
(789,0.35)
(790,0.35)
(791,0.35)
(792,0.35)
(793,0.35)
(794,0.35)
(795,0.35)
(796,0.35)
(797,0.35)
(798,0.35)
(799,0.35)
(800,0.35)
(801,0.35)
(802,0.35)
(803,0.35)
(804,0.35)
(805,0.35)
(806,0.35)
(807,0.35)
(808,0.35)
(809,0.35)
(810,0.35)
(811,0.35)
(812,0.35)
(813,0.35)
(814,0.35)
(815,0.35)
(816,0.35)
(817,0.35)
(818,0.35)
(819,0.35)
(820,0.35)
(821,0.35)
(822,0.35)
(823,0.35)
(824,0.35)
(825,0.35)
(826,0.35)
(827,0.35)
(828,0.35)
(829,0.35)
(830,0.35)
(831,0.35)
(832,0.35)
(833,0.35)
(834,0.35)
(835,0.35)
(836,0.35)
(837,0.35)
(838,0.35)
(839,0.35)
(840,0.35)
(841,0.35)
(842,0.35)
(843,0.35)
(844,0.35)
(845,0.35)
(846,0.35)
(847,0.35)
(848,0.35)
(849,0.35)
(850,0.35)
(851,0.35)
(852,0.35)
(853,0.35)
(854,0.35)
(855,0.35)
(856,0.35)
(857,0.35)
(858,0.35)
(859,0.35)
(860,0.35)
(861,0.35)
(862,0.35)
(863,0.35)
(864,0.35)
(865,0.35)
(866,0.35)
(867,0.35)
(868,0.35)
(869,0.35)
(870,0.35)
(871,0.35)
(872,0.35)
(873,0.35)
(874,0.35)
(875,0.35)
(876,0.35)
(877,0.35)
(878,0.35)
(879,0.35)
(880,0.35)
(881,0.35)
(882,0.35)
(883,0.35)
(884,0.35)
(885,0.35)
(886,0.35)
(887,0.35)
(888,0.35)
(889,0.35)
(890,0.35)
(891,0.35)
(892,0.35)
(893,0.35)
(894,0.35)
(895,0.35)
(896,0.35)
(897,0.35)
(898,0.35)
(899,0.35)
(900,0.35)
			};
		\node  [color=blue] at (rel axis cs:0.47,0.95){Original vs. Reconstructed Feature Similarity};
		\node  [color=red] at (rel axis cs:0.5,0.4){Original vs. Decompiled Feature Average Similarity};
		\end{axis}
\end{tikzpicture}
	\caption{ \label{fig:optim} \small{Feature Transformations: Each data point on the x-axis is a different executable binary sample. Each y-axis value is the cosine similarity between the feature vector extracted from the original source code and the feature vector that tries to \emph{predict} the original features. The average value of these 900 cosine similarity measurements is 0.81, suggesting that decompiled code preserves transformed forms of the original source code features well enough to reconstruct the original source code features.}}
	\end{center}
	\vspace{-0.7cm}
\end{figure}

%% file: 8_discussion.tex
%a)	Discuss results
%i)	default l0, vs optimization level1, vs level2
%ii)	plus other optimization levels
%b)	Future Work
%i)	malware family classification
%(1)	requires a dataset

\section{Discussion} \label{sec:discussion}

% fabs: we say this in limitations, so I'm removing it.

Our experiments are devised for a setting where the programmer is not trying to hide her coding style, and therefore, only basic obfuscation techniques are considered in our experiments. Accordingly, we focus on the general case of executable binary authorship attribution, which is a serious threat to privacy but at the same time an aid for forensic analysis.

We consider two data sets: the GCJ dataset, and a dataset based on GitHub repositories. Using the GitHub dataset, we show that we can perform programmer de-anonymization with executable binary authorship attribution in the wild. We de-anonymize GitHub programmers by using stylistic features obtained from the GCJ dataset. Using the same small set of features, we perform a case study on the leaked hacker forum Nulled.IO and de-anonymize four of its members. The successful de-anonymization of programmers from different sources supports the supposition that, in addition to its other useful properties for scientific analysis of attribution tasks, the GCJ dataset is a valid and useful proxy for real-world authorship attribution tasks. 

%This shows that there is nothing special about the GCJ dataset and it is useful for developing a robust binary code authorship attribution approach with controlled experiments.
%* instead of "there is nothing special about the google code jam dataset" maybe "this supports the supposition that, in addition to its other useful properties for scientific analysis of %attribution tasks, the google code jam dataset is a valid and useful proxy for real-world source and binary attribution tasks"

The advantage of using the GCJ dataset is that we can perform the experiments in a controlled environment where the most distinguishing difference between programmers' solutions is their programming style. Every contestant implements the same functionality, in a limited amount of time while at each round problems are getting more difficult. This provides the opportunity to control the difficulty level of the samples and the skill set of the programmers in the dataset. In source code authorship attribution, programmers who can implement more sophisticated functionality have a more distinct programming style~\cite{caliskan2015anonymizing}. We observe the same pattern in executable binary samples and gain some software engineering insights by analyzing stylistic properties of executable binaries. In contrast to GCJ, GitHub and Nulled.IO offer noisy samples. However, our results show that we can de-anonymize programmers with high accuracy as long as enough training data is available.

Previous work shows that coding style is quite prevalent in source code. We were surprised to find that it is also preserved to a great degree in compiled source code. Coding style is not just the use of particular syntactical constructs but also the AST flows, AST combinations, and preferred types of operations. Consequently, these patterns manifest in the binary and form a coding fingerprint for each author. We can de-anonymize programmers from compiled source code with great accuracy, and furthermore, we can de-anonymize programmers from source code compiled with optimization or after obfuscation. In our experiments, we see that even though basic obfuscation, optimization, or stripping symbols transforms executable binaries more than plain compilation, stylistic features are still preserved to a large degree. Such methods are not sufficient on their own to protect programmers from de-anonymization attacks.

In scenarios where authorship attribution is challenging, an analyst or adversary could apply relaxed attribution to find a suspect set of $n$ authors, instead of a direct {\em top--1} classification. In {\em top--10} attribution, the chances of having the original author within the returned set of $10$ authors approaches 100\%. Once the suspect set size is reduced to $10$ from $hundreds$, the analyst or adversary could adhere to content based dynamic approaches and reverse engineering to identify the author of the executable binary sample. However, our experiments in these cases are performed using the information-gain features determined from the unoptimized case with symbol tables intact. Future work that customizes the dimensionality reduction step for these cases (for example, removing features from the trees that are no longer relevant) may be able to improve upon these numbers, especially since dimensionality reduction was able to provide such a large boost in the unoptimized case.

Even though executable binaries look cryptic and difficult to analyze, we can still extract many useful features from them. We extract features from disassembly, control flow graphs, and also decompiled code to identify features relevant to only programming style. After dimensionality reduction, we see that each of the feature spaces provides programmer style information. The initial development feature set contains a total of 705,000 features for 900 executable binary samples of 100 authors. Approximately 50 features from abstract syntax trees and assembly instructions suffice to capture enough key information about coding style to enable robust authorship attribution. We see that the reduced set of features are valid in different datasets with different programmers, including optimized or obfuscated programmers. Also, the reduced feature set is helpful in scaling up the programmer de-anonymization approach. While we can identify 100 programmers with 96\% accuracy, we can de-anonymize 600 programmers with 83\% accuracy using the same reduced set of features. 83\% is a very high number for such a challenging task where the random chance of correctly identifying an author is 0.17\%. 

\section{Limitations} \label{sec:limitations}

Our experiments suggest that our method is able to assist in de-anonymizing a much larger set of programmers with significantly higher accuracy than state-of-the-art approaches. However, there are also assumptions that underlie the validity of our experiments as well as inherent limitations of our method which we discuss in the following paragraphs. First, we assume that our ground truth is correct, but in reality programs in GCJ or on GitHub might be written by programmers other than the stated programmer, or by multiple programmers. Such a ground truth problem would cause the classifier to train on noisy models which would lead to lower de-anonymization accuracy and a noisy representation of programming style. Second, many source code samples from GCJ contestants cannot be compiled. Consequently, we perform evaluation only on the subset of samples which can be compiled. This has two effects: first, we are performing attribution with fewer executable binary samples than the number of available source code samples. This is a limitation for our experiments but it is not a limitation for an attacker who first gets access to the executable binary instead of the source code. If the attacker gets access to the source code instead, she could perform regular source code authorship attribution. Second, we must assume that whether or not a code sample can be compiled does not correlate with the ease of attribution for that sample.
Third, we mainly focus on C/C++ code compiled (except Nulled.IO samples) using the GNU compiler \code{gcc} in this work, and assume that the executable binary format is the Executable and Linking Format. This is important to note as dynamic symbols are typically present in ELF binary files even after stripping of symbols, which may ease the attribution task relative to other executable binary formats that may not contain this information. We defer an in depth investigation of the impact that other compilers, languages, and binary formats might have on the attribution task to future work.

Finally, while we show that our method is capable of dealing with simple binary obfuscation techniques, we do not consider binaries that are heavily obfuscated to hinder reverse engineering. While simple systems, such as packers \cite{upx} or encryption stubs that merely restore the original executable binary into memory during execution may be analyzed by simply recovering the unpacked or decrypted executable binary from memory, more complex approaches are becoming increasingly commonplace. A wide range of anti-forensic techniques exist \cite{unpacker}, including methods that are designed specifically to prevent easy access to the original bytecode in memory via such techniques as modifying the process environment block or triggering decryption on the fly via guard pages. Other techniques such as virtualization \cite{oreans} transform the original bytecode to emulated bytecode running on virtual machines, making decompilation both labor-intensive and error-prone. Finally, the use of specialized compilers that lack decompilers and produce nonstandard machine code (see \cite{movfuscator} for an extreme but illustrative example) may likewise hinder our approach, particularly if the compiler is not available and cannot be fingerprinted. We leave the examination of these techniques, both with respect to their impact on authorship attribution and to possible mitigations, to future work.

%~\citep{MarMarGua15}

%% file: 9_conclusion.tex
\section{Conclusion}
\label{sec:conclusion}

De-anonymizing programmers has direct implications for privacy and anonymity. The ability to attribute authorship to anonymous executable binaries has applications in software forensics, and is an immediate concern for programmers that would like to remain anonymous. We show that coding style is preserved in compilation, contrary to the belief that compilation wipes away stylistic properties. We de-anonymize 100 programmers from their executable binaries with 96\% accuracy, and 600 programmers with 83\% accuracy. Moreover, we show that we can de-anonymize GitHub developers or hacker forum members with high accuracy. Our work, while significantly improving the limited approaches in programmer de-anonymization, presents new methods to de-anonymize programmers in the wild from challenging real-world samples.

We discover a small set of features that effectively represent coding style in executable binaries. We obtain this precise representation of coding style via two different disassemblers, control flow graphs, and a decompiler. With this comprehensive representation, we are able to re-identify GitHub authors from their executable binary samples in the wild, where we reach an accuracy of 65\% for 50 programmers, even though these samples are noisy and products of collaborative efforts.

Programmer style is embedded in executable binary to a surprising degree, even when it is obfuscated, generated with aggressive compiler optimizations, or symbols are stripped. Compilation, binary obfuscation, optimization, and stripping of symbols reduce the accuracy of stylistic analysis but are not effective in anonymizing coding style.

In future work, we plan to investigate snippet and function level stylistic information to de-anonymize multiple authors of collaboratively generated binaries. We also defer the analysis of highly sophisticated compilation and obfuscation methods to future work. Nevertheless, we show that identifying stylistic information is prevalent in real-world settings and accordingly developers cannot assume to be anonymous unless they take extreme precautions as a countermeasure. Examples to possible countermeasures include a combination of randomized coding style, different programming language usage, and employment of indeterministic set of obfuscation methods. Since incorporating different languages or obfuscation methods is not always practical, especially in open source software, our future work would focus on completely stripping stylistic information from binaries to render them anonymous.

We also plan to look at different real-world executable binary authorship attribution cases, such as identifying authors of malware, which go through a mixture of sophisticated obfuscation methods by combining polymorphism and encryption. Our results so far suggest that while stylistic analysis is unlikely to provide a ``smoking gun'' in the malware case, it may contribute significantly to attribution efforts.

% fabs: We can add this, but it doesn't seem as important as the rest.

Moreover, we show that attribution is sometimes possible with only small amounts of training binaries, however, having more binaries for training helps significantly. In addition, we observe that advanced programmers (as measured by progression in the GCJ contest) can be attributed more easily than their less skilled peers. Our results present a privacy threat for people who would like to release binaries anonymously.

% rich: the removed part seems redundant; tried to rework a bit
% fabs: removed because we've said this at the beginning of the paragraph.
%\aylin{multi author baa, trace a snippet in a binary, function level baa\\}